\documentclass[twocolumn]{aastex63}
\usepackage{graphicx}
\usepackage{amssymb}
\usepackage{amsmath}
\usepackage{url}
\usepackage{color, colortbl}
\usepackage{textgreek}
\usepackage{xcolor}

\usepackage{color, colortbl}
\definecolor{Gray}{gray}{0.75}

\usepackage{xcolor}
\usepackage{soul}

\accepted{for publication on January 25, 2022}

\begin{document}

\title{Revisiting Kepler Transiting Systems: Unvetting Planets and Constraining Relationships among Harmonics in Phase Curves}

\author[0000-0002-8052-3893]{Prajwal Niraula}
\affiliation{Department of Earth, Atmospheric and Planetary Sciences, MIT, 77 Massachusetts Avenue, Cambridge, MA 02139, USA; pniraula@mit.edu}

\author[0000-0002-1836-3120]{Avi Shporer}
\affiliation{Department of Physics and Kavli Institute for Astrophysics and  Space  Research,  Massachusetts  Institute  of  Technology, Cambridge, MA 02139, USA}

\author[0000-0001-9665-8429]{Ian Wong}
\altaffiliation{NASA Postdoctoral Program Fellow}
\affiliation{NASA Goddard Space Flight Center, 8800 Greenbelt Rd, Greenbelt, MD 20771, USA}

\author[0000-0001-9665-8429]{Julien de Wit}
\affiliation{Department of Earth, Atmospheric and Planetary Sciences, MIT, 77 Massachusetts Avenue, Cambridge, MA 02139, USA}

\begin{abstract}
Space-based photometric missions widely use statistical validation tools for vetting transiting planetary candidates, particularly when other traditional methods of planet confirmation are unviable. In this paper, we refute the planetary nature of three previously validated planets --- Kepler-854~b, Kepler-840~b, and Kepler-699~b --- and possibly a fourth, Kepler-747~b, using updated stellar parameters from Gaia and phase-curve analysis. In all four cases, the inferred physical radii rule out their planetary nature given the stellar radiation the companions receive. For Kepler-854~b, the mass derived from the host star's ellipsoidal variation, which had not been part of the original vetting procedure, similarly points to a non-planetary value. To contextualize our understanding of the phase curve for stellar mass companions in particular and extend our understanding of high-order harmonics, we examine Kepler eclipsing binaries with periods between 1.5 and 10 days. Using a sample of 20 systems, we report a strong power-law relation between the second cosine harmonic of the phase-curve signal and the higher cosine harmonics, which supports the hypothesis that those signals arise from the tidal interaction between the binary components. We find that the ratio between the second and third-harmonic amplitudes is $2.24 \pm 0.48$, in good agreement with the expected value of 2.4 from the classical formalism for the ellipsoidal distortion. 
\end{abstract}


\section{Introduction} 
\label{sec:intro}

Space-based searches for transiting planets, including CoRoT \citep{auvergne2009}, Kepler \citep{borucki2010}, K2 \citep{howell2014}, and TESS \citep{ricker2014}, have detected thousands of transiting planet candidates around faint host stars. The observational resources required to confirm or refute the planetary nature of each candidate are prohibitively large, especially in the case of the faintest targets. The necessity to vet the numerous candidates has led to the development of statistical validation techniques \citep[e.g.,][]{torres2011, torres2015, torres2017, morton2011, morton2012, diaz2014, morton2016}. Such frameworks provide estimates of the probability that a candidate is a false positive using a combination of stellar parameters and transit-derived properties. A candidate is declared as a validated planet if the false positive probability is below a certain threshold (usually 0.01 or 0.001, e.g. \citealt{crossfield2016}). Since the Kepler era, these methods have been widely used for space-based photometric missions, including K2 \citep{montet2015, crossfield2016, mayo2018, livingston2018, deleon2021} and TESS \citep{giacalone2021}.
 
Despite the popularity and broad applicability of statistical validation methods, there are cases where statistically validated planets have been subsequently shown not to be bonafide planets through more resource-intensive follow-up observations, such as radial velocity monitoring \citep[e.g.,][]{shporer2017vespa}. Similarly, revised stellar parameters, notably those provided by the Gaia mission \citep{GaiaDR2, gaiaEDR3}, can reveal misclassified planets as we will be demonstrating in this paper.

Beyond radial velocity (RV) follow-up and revised stellar parameters, phase curves, which denote photometric modulations that are synchronous with the orbital period, provide a powerful tool for gaining insights into the mutual gravitational and radiative interactions between objects in binary systems. The three primary astrophysical phenomena driving these modulations are (1) ellipsoidal variation originating from the tidal interaction, (2) beaming (or Doppler boosting) due to relativistic Doppler effects, and (3) mutual illumination due to the mutual heating of the components' atmospheres. In the past, phase curves have been used to validate exoplanets, such as Kepler-41~b \citep{quintana2013} and Kepler-78~b \citep{ojeda2013}. For short-period candidates (typically shorter than 5 days), the phase curve can provide an independent estimate of the companion mass and complement other planet characterization techniques without requiring further observations \citep[e.g.,][]{faigler2011,shporer2017}.

In this paper, we reassess the planetary nature of Kepler companions in light of updated stellar catalogs and then consider how phase-curve photometry can be used to mitigate the occurrence of misclassified companions. We find that three of them, and possibly a fourth, cannot be planets or brown dwarfs and are thus stellar objects that have been misclassified as planets. Then, we use a sample of Kepler eclipsing stellar binaries to show that the higher-order harmonic seen in the phase curve of one of the misclassified Kepler planets is likely to be a manifestation of the host star's tidal distortion. This paper is organized as follows. We describe the methods used in vetting the planetary nature of companions in Section~\ref{sec:Tools}. In Section~\ref{sec:KeplerStars}, we apply these techniques to argue against the planetary nature of four previously validated Kepler planets. In Section~\ref{sec:dataanal}, we extend our study of orbital phase curves to a sample of Kepler eclipsing stellar binaries to investigate the higher-harmonic phase-curve modulations. Finally, a summary of our main findings is given in Section~\ref{sec:conclusion}.

\section{Revisiting Validated Kepler Planets} 
\label{sec:Tools}

\begin{figure*}[t]
\begin{tabular}{lr}
     \includegraphics[width=0.48\textwidth, trim={0.5 0.5cm 0 0.5cm},clip ]{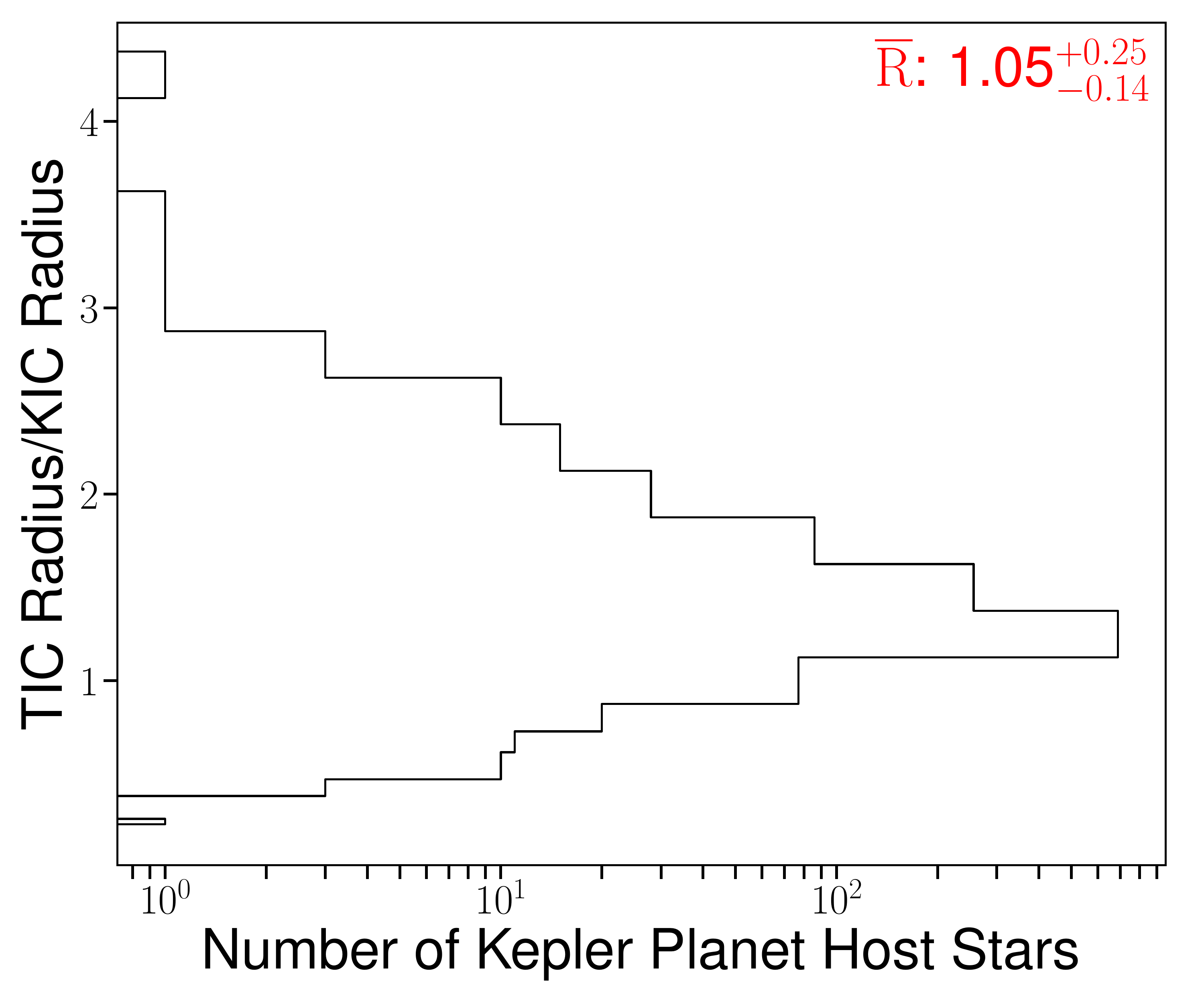} 
     & \includegraphics[width=0.48\textwidth, trim={0.5 0.5cm 0 0.5cm},clip]{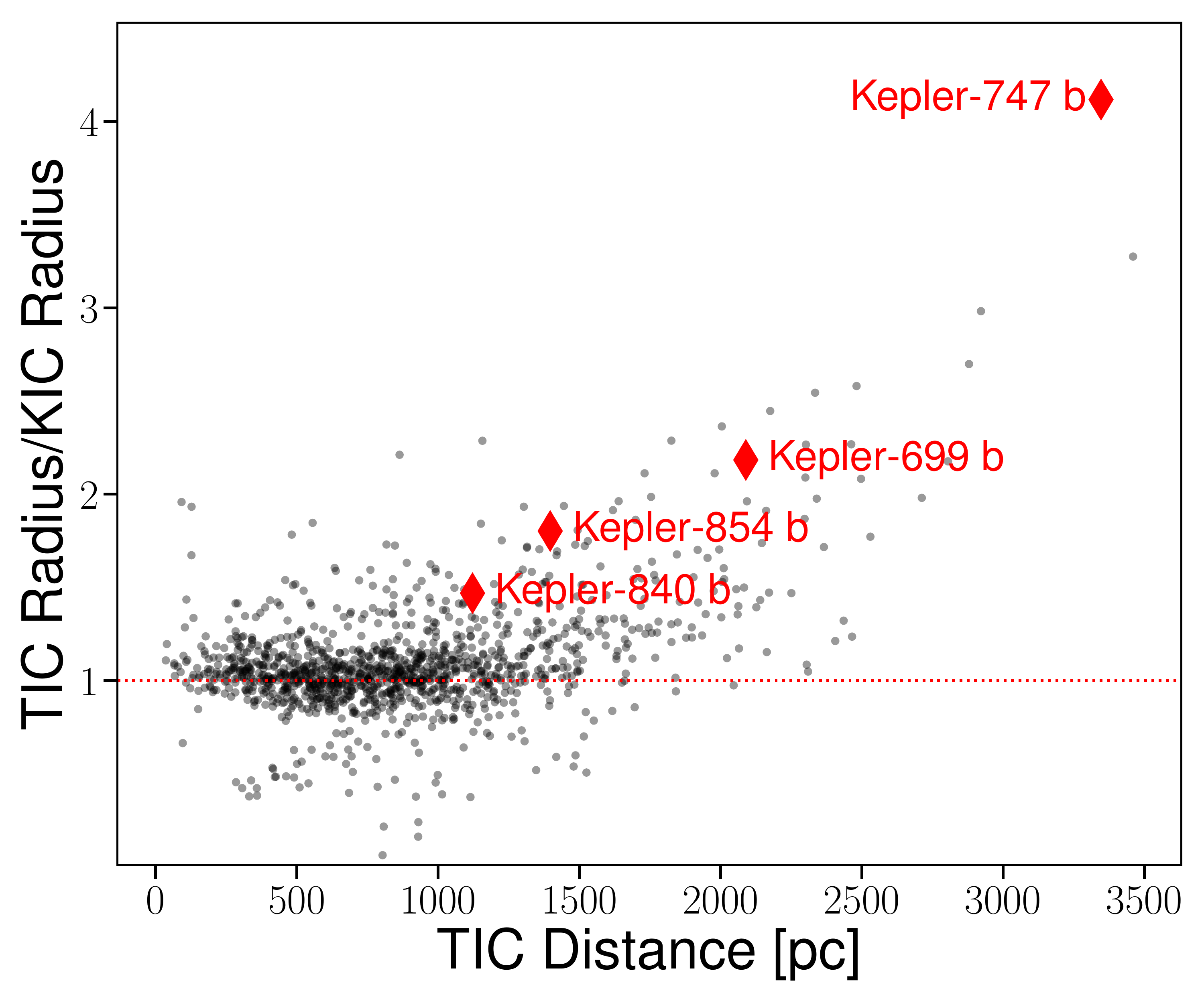} \\
     \includegraphics[width=0.48\textwidth, trim={0.5 0.5cm 0 0.5cm},clip ]{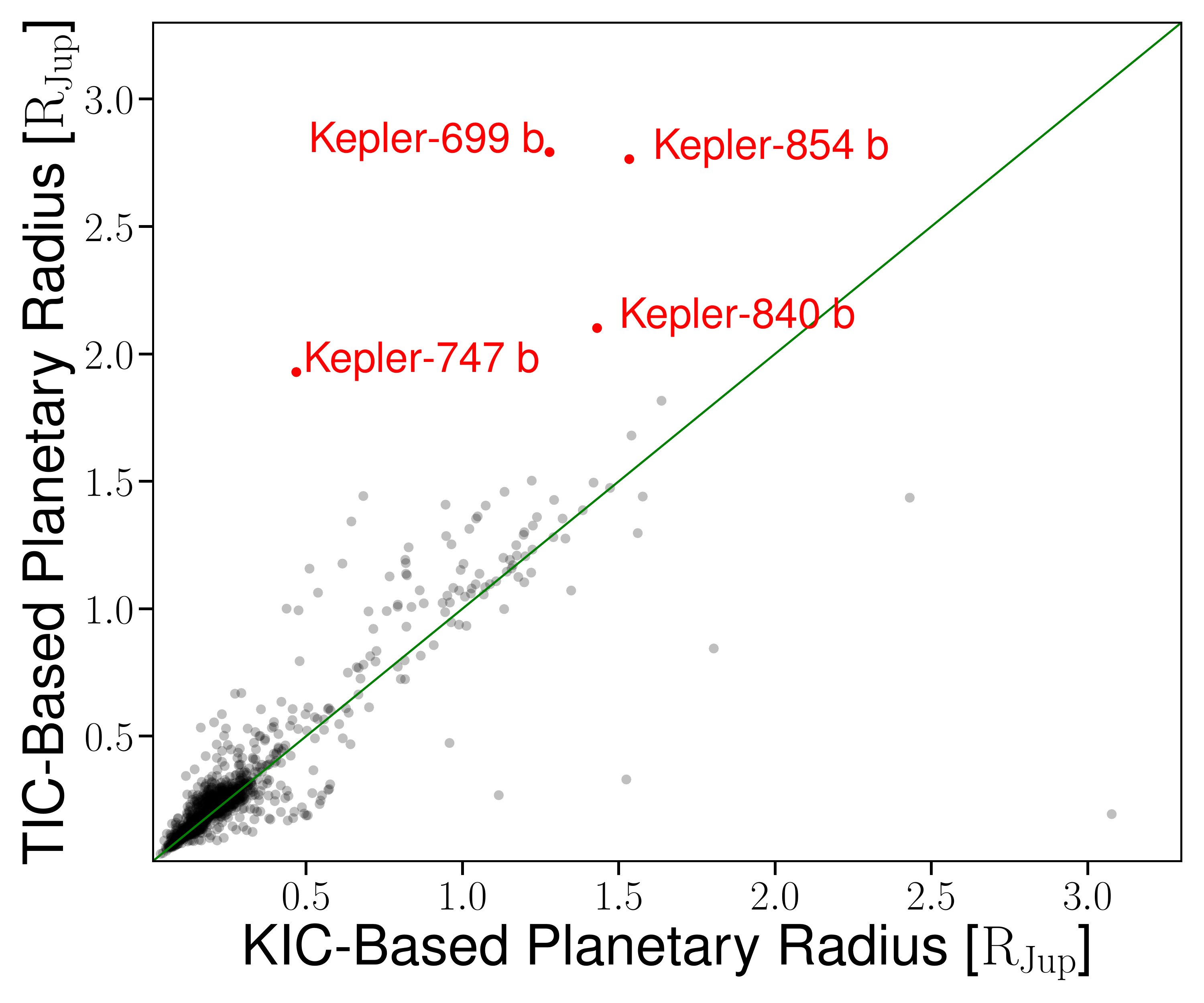} 
     & \includegraphics[width=0.48\textwidth, trim={0.5 0.5cm 0 0.5cm},clip]{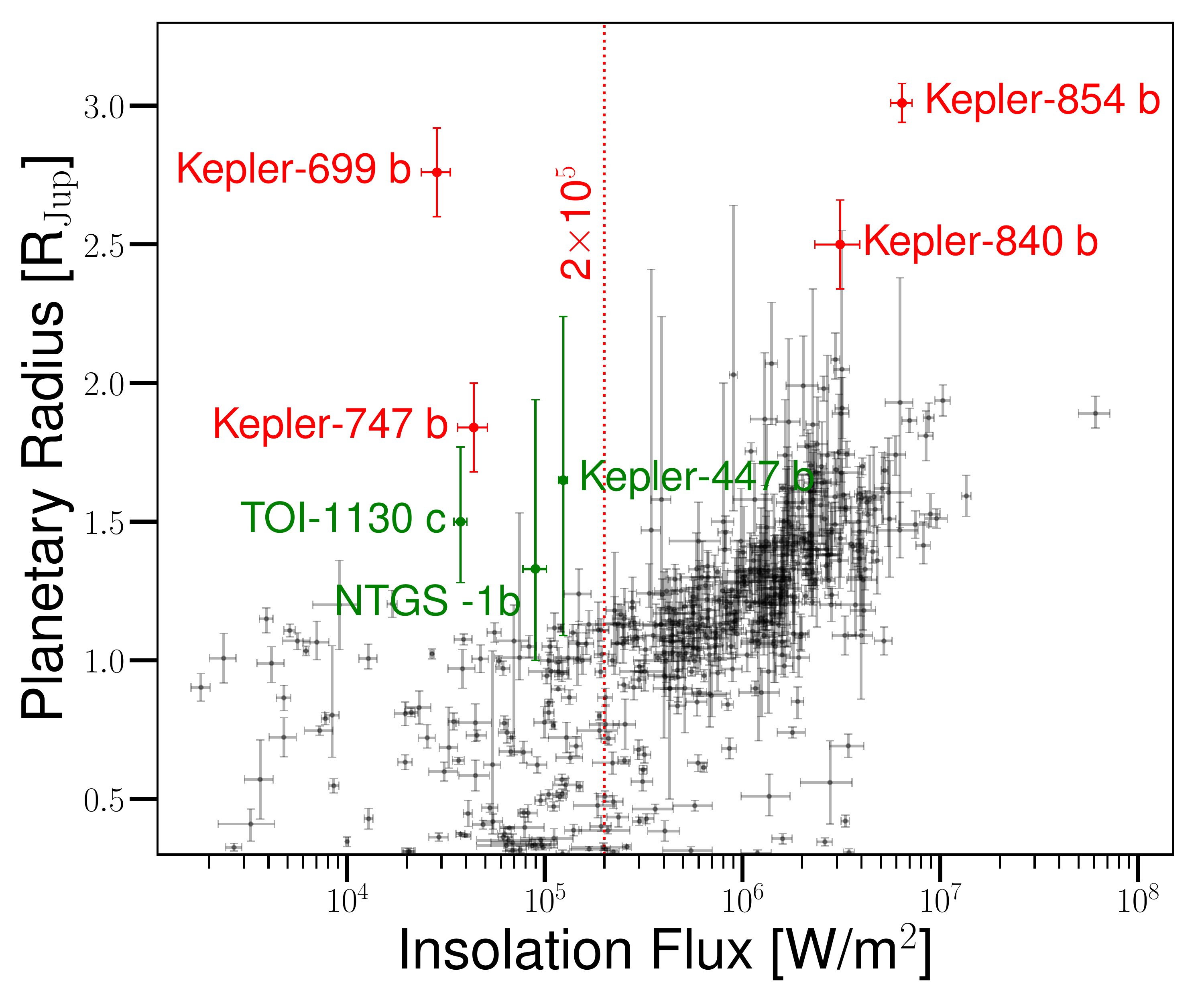} \\
\end{tabular}
\caption{\label{fig:RadiusRatio} Top left: histogram of the ratio $\bar{R}$ between the TIC-based and KIC-based stellar radius for confirmed Kepler planet-hosting stars. About 3.5\% of hosts had their radius over- or underestimated by a factor larger than 2 in the KIC.  Top right: TIC-to-KIC radius ratio as a function of the distance listed in the TIC. Cases with larger ratios typically correspond to targets that lie at heliocentric distances greater than 1,250 pc. Bottom left: a direct comparison of the TIC-based planetary radii versus the KIC-based planetary radii. The 1:1 line is shown in green for guidance. The four previously validated candidates with the largest TIC radius values are marked in red. Bottom right: plot of planet radius vs. insolation calculated assuming a circular orbit, which shows the radius inflation trend beyond an insolation flux of around $2\times10^5$ W/m$^2$  \citep{demory2011}. The data were obtained from the NASA Exoplanet Archive in November 2021. It is evident that four of our targets --- Kepler-854~b, Kepler-840~b,  Kepler-699~b, and Kepler-747~b --- have discrepant radii, especially the latter two, given their insolation levels. This indicates that these companions are not planetary in nature. Kepler-447~b, NGTS-1~b, and TOI-1130~c, which lie near Kepler-747~b in this plot, are grazing planets with larger than typical radius uncertainties.}
\end{figure*}

Using the Kepler photometric data along with updated stellar parameters can help us identify some of the false positives hidden among the current planetary population. Better constraints on the distance from astrometric missions such as Gaia \citep{GaiaDR2, gaiaEDR3} provide better estimates of the physical radius of the stars, thereby leading to more accurate physical radii of the transiting companions. Meanwhile, phase curves, which typically are not used for vetting planetary candidates, enable us to infer the mass of the companion via the characteristic modulations that arise from the mutual gravitational interaction. Usually, the amplitude of the ellipsoidal variation in optical wavelengths does not exceed 200 ppm in a planet--star system \citep[see][]{millholland2017}, allowing us to rule out some of the obvious false positives. Further analysis of the phase curve, especially the amplitudes at the first harmonic of the orbital frequency, can indicate if the transiting companion is self-luminous (i.e., a star) or a nonluminous planet. 

\subsection{Updated Stellar Parameters}
The parameters of transiting companions, such as radius, are often derived from measured quantities that are defined relative to the host star's properties. Therefore, any bias in the stellar parameters will propagate directly as a bias in the companion's properties. We searched for substantial changes in the inferred radii of Kepler planets that are caused by revised stellar parameters. To that end, we first considered the differences in stellar radius of the confirmed Kepler planet hosts between the Kepler Input Catalog (KIC; original catalog) \citep{brown2011} and Version 8 of the TESS Input Catalog (TIC v8.2; revised catalog) \citep{stassun2019}. As seen in the top-left panel of Figure~\ref{fig:RadiusRatio}, we found that the majority of host stars have comparable stellar radius values in the two catalogs. However, $\sim$3.5\% of the stars show radius discrepancies of larger than a factor of two. These differences are primarily driven by the updated heliocentric distances provided by the Gaia DR2 parallax measurements \citep{GaiaDR2} that were used to estimate the TIC stellar radii.  The histogram also reveals an asymmetric distribution, with the deviation strongly skewed toward larger ratios (+0.25 versus $-$0.14). For most objects beyond 1,250 pc, the stellar radii have been revised to larger values owing to the previous underestimation of their distance (see the top-right panel of Figure~\ref{fig:RadiusRatio}).

For the next step of our assessment of Kepler planetary systems, we took the published planet--star radius ratio for each planet-hosting system and calculated the planet radius, using both KIC or TIC value for the host star's radius. The bottom-left panel of Figure~\ref{fig:RadiusRatio} shows the comparison between the TIC- and KIC-based planet radii. Focusing on the upper half of the graph, four putative planets clearly stand out as having significantly larger TIC-based radii, and are therefore suspect as possible misclassified objects. For these systems, we performed independent fits to the host star's stellar energy distribution (SED) using the \texttt{isochrone} package \citep{morton2015}. In this analysis, we used publicly available broadband photometric measurements from the AllWISE source catalog ($B$, $J$, $H$, $K_{s}$, $W1$, $W2$, $W3$, $W4$; \citealt{cutri2013}), alongside the Gaia Early Data Release 3 (EDR3) parallax \citep{gaiaEDR3}. As a separate check for consistency, we compared the density of the host star obtained from our SED analysis to the density derived from the Kepler transit fit assuming a circular orbit (described, for example, in Equation~30 of  \citealt{winn2010}).

\subsection{Phase-curve Model} 
\label{sec:model}
Let $\Phi$ represent the orbital phase from the mid-transit time (i.e., inferior conjunction). We can mathematically express $\Phi$ as follows:

\begin{equation}
    \Phi = ([t - T_0]~\mathrm{mod}~P)/P.
\end{equation}
Here, $T_0$ is the time of mid-transit, and $P$ is the orbital period. In our phase-curve analysis, we considered up to five harmonics of the orbital phase, with both sine and cosine terms contribute to the phase curve flux ($\mathcal{L}_{\rm{PC}}$):

\begin{equation}
    \mathcal{L}_{\rm{PC}} = 1 + \sum_{k=1}^{5}\left( A_k \sin 2\pi k \Phi + B_k \cos 2\pi k \Phi\right).
\label{eqn:phasemodel}
\end{equation}
Note that in this formulation all components are independent of each other. Many of these components can be directly tied to the various astrophysical phenomena that occur in binary systems. The first cosine harmonic ($B_1$) originates from the mutual illumination of the stars, whereas the ellipsoidal effect dominates the second ($B_2$) and higher ($B_3$, $B_4$) cosine harmonics \citep{kopal1959}. The beaming effect itself manifests as the first sine term ($A_1$). While the higher-order sine amplitudes are not expected from theoretical modeling of binary-system phase curves, they have been observed for several systems in the past \citep[e.g.][]{rappaport2015, WongKOI964_2020}, and therefore we include them for completeness. 

For binary systems in which both components have significant luminosity, it is important to account for the contributions from both stars. As the orbital phase of the secondary is shifted by 1/2 relative to the primary, all of the odd harmonics for both sine and cosine components ($A_1$, $A_3$, $A_5$, $B_1$, $B_3$, and $B_5$) are subtractive, whereas the even harmonics ($A_2$, $A_4$, $B_2$, and $B_4$) are additive. As a result, if the contribution from the secondary object dominates, the predicted signals calculated when assuming the primary star is brighter than the secondary may not only show a mismatch with the measurements but could even end up with a reversed sign altogether. In the following subsections, we describe the major phase-curve components.

\subsubsection{Ellipsoidal Variation}
\label{sec:ellipsoidal}
The ellipsoidal variation arises from the mutual gravitational interaction between the two binary components. The primary physical processes that drive this modulation are the equilibrium tidal bulge raised on the primary star by the companion and the limb- and gravity-darkening of the distended atmosphere. The analytical formulation describing ellipsoidal variation in stellar atmospheres was first put forth by \citet{kopal1959}, with updated mathematical expressions of the corresponding phase-curve amplitudes provided more recently in \citet{morris1985} and \citet{morris1993}. Generally, the ellipsoidal  induced flux variation ($\mathcal{L}_{\rm{E}}$) in the phase curve is expressed as a series of cosine terms: 

\begin{equation}
    \mathcal{L}_{\rm{E}} = 1 + \sum\limits_{k=1}^{5} B_k \cdot \cos (2 \pi k \Phi),
\end{equation}
where $B_k$ are coefficients for the different ellipsoidal harmonics. The classical analytical formulation of ellipsoidal variation provides mathematical expressions for the first four harmonics \citep[e.g.,][]{morris1985}: 

\begin{align}
    B_1 &= \frac{3 q (4 \sin i - 5 \sin^3 i) }{16(a/R_1)^4}  \cdot \mathbb{X}_3 \cdot \mathbb{G}_3, \notag\\
     B_2 &= -\frac{3 q \sin^2 i}{4(a/R_1)^3}\cdot \mathbb{X}_2 \cdot \mathbb{G}_2, \notag \\
    &\quad - \frac{5q}{48(a/R_1)^5}\cdot(6 \sin^2 i - 7 \sin^4 i) \cdot \mathbb{X}_4 \cdot \mathbb{G}_4,  \notag\\
    B_3 &= -\frac{5 q \sin^3{i}}{16(a/R_1)^4} \cdot \mathbb{X}_3 \cdot \mathbb{G}_3, \notag\\
    B_4 &= \frac{35 q \sin ^4 i}{192(a/R_1)^5} \cdot \mathbb{X}_4\cdot\mathbb{G}_4. 
\label{eqn:ellipsoidal_var_eqn}    
\end{align}

In the above equations, $q$ denotes the mass ratio between the binary components, $i$ is the orbital inclination, and $a/R_1$ is the ratio between the orbital semimajor axis and the radius of the primary star. The terms represented by $\mathbb{X}$ and $\mathbb{G}$ contain the contributions from limb- and gravity-darkening, respectively. It has been standard practice in the literature to model ellipsoidal variation assuming linear limb-darkening models. In that case, the contribution of limb- and gravity-darkening can be written as:

\begin{equation}
\begin{split}
   \mathbb{X}_2 &= \frac{15+u}{5(3-u)}, \quad \mathbb{G}_2 = 1+g,  \\
    \mathbb{X}_3 &= \frac{5u}{2(3-u)}, \quad \mathbb{G}_3 = 2+g,\\
    \mathbb{X}_4 &= \frac{9(1-u)}{4(3-u)}, \quad \mathbb{G}_4 = 3+g,\\
\label{eqn:limbdarkening}    
\end{split}
\end{equation}
where $u$ is the linear limb-darkening parameter, and $g$ is the gravity-darkening parameter. For both linear limb-darkening and gravity-darkening parameter values, the values can be interpolated using the relevant tabulated parameters provided in \citet{claret2011}. The precision of these coefficients are estimated through Monte Carlo sampling using the uncertainties on the reported stellar parameters.

Combining the contributions from both components in the binary system, we arrive at the full mathematical expression of the leading-term amplitude of the ellipsoidal variation (i.e., at the second harmonic of the cosine); here the limb- and gravity-darkening terms are not included (they are set to one), for simplicity:

\begin{equation}
\begin{split}
B_2 &\approx -\frac{3 q \sin^2 i}{4(a/R_1)^3} -f_2\cdot\frac{3 \sin^2 i}{4q(a/R_2)^3} \\
&= -\frac{3q \sin^2 i}{4 (a/R_1)^3}\left(1 + f_2\cdot\frac{(R_2/R_1)^3}{q^2} \right).\\
\end{split}
 \end{equation}
Here, $R_2$ represents the radius of the secondary, and $f_2$ denotes the relative luminosity (i.e., flux ratio) of the secondary in the observed bandpass.

For the third-harmonic amplitude, which is typically the second-order term in the description of the photometric ellipsoidal variation, we can combine the two contributions analogously:
 
\begin{equation}
\begin{split}
B_3 &\approx -\frac{5 q \sin^3{i}}{16(a/R_1)^4} + f_2\cdot \frac{5 \sin^3{i}}{16q(a/R_2)^4} \\
&= -\frac{5q \sin^3{i}}{16 (a/R_1)^4} \left(1 - f_2\cdot  \frac{(R_2/R_1)^4}{q^2} \right).
\end{split}
\end{equation}
Now, taking the ratio between these two amplitudes, we obtain

\begin{equation}
\label{eq:b2b3}
\left|\frac{B_2}{B_3}\right| = \frac{12 (a/R_1)}{5 \sin i} \cdot \left(\frac{q^2 + f_2\cdot(R_2/R_1)^3}{q^2 - f_2\cdot(R_2/R_1)^4}\right).
\end{equation}

The contribution of the secondary, as seen in the expression above, depends both on the radius ratio $R_2/R_1$ and the flux ratio $f_2$. Our ideal targets contain secondaries that are significantly smaller than the primary (at least by a factor of 2) and contribute negligible flux (in the optical) compared to that of the primary star. In this scenario, both the nominator and the denominator in the parenthesis in Equation~\ref{eq:b2b3} can be approximated as $q^2$ and thus cancel out, making the ratio $|B_2/B_3|$ independent of $q$.  Reintroducing the limb- and gravity-darkening terms of the primary star for systems where the secondary is small and faint, we arrive at the approximate expression

\begin{equation}
\label{eqn:B2B3Ratio}
\left|\frac{B_2}{B_3}\right| \approx \frac{12 (a/R_1)}{5 \sin i} \cdot \left(\frac{\mathbb{X}_2 \cdot \mathbb{G}_2 }{\mathbb{X}_3 \cdot \mathbb{G}_3}\right) \ . 
\end{equation}

We note that the limb- and gravity-darkening terms at the second and third harmonics can differ by more than a factor of 2, so careful modeling of their effects is crucial for studying the precise relationship between the ratio of $B_2$ and $B_3$. Besides these terms, all of the parameters can be determined from the fit, and so the above equation allows us to test the classical formulation of \citet{kopal1959} without needing to know the mass ratio.

\subsubsection{Beaming Effect} 
\label{sec:beaming}

Beaming, otherwise known as Doppler boosting, is a relativistic effect that is caused by the line-of-sight acceleration of the two binary components around the common center of mass. For circular systems, this effect manifests itself as a photometric modulation at the first harmonic of the sine of the orbital phase and mirrors the shape of the radial velocity curve, albeit with the opposite sign \citep{loeb2003}:

\begin{align}
    A_1 &= \alpha_1 \frac{K_{\mathrm{RV},1}}{c} - \alpha_2 f_2 \frac{K_{\mathrm{RV},2}}{c} \notag\\
    &= \left(\alpha_1 M_2  - \alpha_2 f_2 M_1  \right) \frac{\sin i}{c}\left( \frac{2\pi G}{P} \frac{1}{(M_1+M_2)^2}\right)^{1/3}.
\label{eqn:DopplerBeaming}
\end{align}
Here, $M_1$ and $M_2$ are the masses of the two binary components. The coefficients of the beaming effect $\alpha_i$ are given by

\begin{equation}
    \alpha_i = 3 - \left\langle \frac{d \log F_\nu}{d \nu}\right\rangle,
\end{equation}
where $F_\nu$ is the emission spectrum of the star in frequency units integrated over the Kepler bandpass. In our analysis, we used PHOENIX stellar models \citep{husser2013} to estimate the expected amplitude of the beaming effect. If both of the stars are on the main sequence, the luminosity depends strongly on mass ($L\propto M^\gamma$, where $\gamma$ is close to 4). It follows that the theoretical amplitude in  Equation~\eqref{eqn:DopplerBeaming} should always be positive in the case of a main-sequence binary. Likewise, in a star--planet system, $f_2$ is negligible, so $A_1$ is once again expected to be positive.

\subsubsection{Mutual Illumination} 
\label{sec:illu}
While stars are poor reflectors overall, the temperature difference between the primary and secondary as well as their large surface areas can lead to non-negligible mutual illumination effects.  This effect is interpreted as heating of the stellar atmosphere, which in turn raises the local brightness and produces photometric modulations as the two components orbit each other.  \citet{kopal1959} showed that the mutual illumination is manifested in the phase curve as a series of cosine terms, with the first harmonic ($B_1$) being the dominant term. Depending on the surface brightness ratio, the value of $B_1$ stemming from mutual illumination is expected to be negative for typical main-sequence binaries, but maybe positive for systems where the secondary has a significantly higher luminosity than the primary, such as in the case of binaries consisting of a hot white dwarf orbiting a main sequence star \citep[e.g.,][]{WongKOI964_2020}.

\section{Kepler Planets Reclassified as Stellar Companions}
\label{sec:KeplerStars}

\begin{deluxetable*}{rccccccc}
\tablecaption{System Parameters for Misclassified Planets}
\tablehead{
\colhead{Parameter} & \colhead{Unit} & \colhead{Kepler-854~b} &  \colhead{Kepler-840~b}  &  \colhead{Kepler-699~b} & \colhead{Kepler-747~b} 
}
\startdata
\multicolumn{3}{l}{\quad Stellar parameters} \\
KIC  & \dots  & 7532973 & 11517719 &   6061119 & 5780460\\
Kepler Mag  & \dots & 13.48  & 14.15   &  15.48 & 15.70 \\
$R_1$ (M16)$^{\dagger}$ &  $R_\odot$ &  1.25$^{+0.51}_{-0.17}$ & 1.06$^{+0.24}_{-0.12}$ &  0.780$^{+0.077}_{-0.067}$ & 0.780$\pm$0.042  \\
$R_1$ (isochrones) &  $R_\odot$ &  2.56(5) & 1.73(11) &  1.71(10) & 3.20(29)  \\
$M_1$ (isochrones) & $M_\odot$ & 0.84(4) &  1.20(10) & 0.92(5) & 0.78(13) \\ 
$T_1$ (isochrones) & K & 6030(160) & 6220(230) &  5530(230) & 5420(210) \\
$\log g$ (isochrones) & \dots & 3.53(3) & 4.04(6) &  3.93(6) & 3.31(9) \\
Age (isochrones) & Gyr & 10.00(10) & 9.63(13) &  10.06(11) & 6.70(23) \\
Fe/H (isochrones) & dex & $-$2.26(21) & $-$0.18(0.22) & 0.31(22) & $-$2.60(38) \\
\hline
\multicolumn{3}{l}{\quad Orbital, transit, and secondary eclipse parameters}\\
$P$ & day &  2.14460946(25) & 2.49578284(21)  & 27.8075643(40)  &  35.61757(4)  \\ 
$T_0$ ($\mathrm{BJD}_{\mathrm{TDB}}-2450000$) & day & 4966.992805(78) & 4968.33697(15)  & 5019.71363(11) & 5030.03642(87) \\ 
$R_2/R_1$ & \dots & 0.1212(19) & 0.1449(22)  & 0.1622(15) & 0.05771(75) \\
$b$ & \dots & 0.771(32) & 0.23(15) & 0.7833(82) &  0.22(16)\\
$a/R_1$ & \dots &  3.42(12) &  5.21(21) & 43.21(46) & 33.5(13)  \\
$u_1$ & \dots & 0.30(26) & 0.32(13) & 0.46(28) & 0.415$\pm$(97) \\
$u_2$ & \dots & 0.21(35) & 0.16(29) & 0.13(34) & 0.32(0.19)\\
\hline
\multicolumn{3}{l}{\quad Phase-curve parameters} \\
$A_1$  & ppm & 125(24) & $\dots$ &  $\dots$& $\dots$\\ 
$B_1$ & ppm & $-$532(35) & $\dots$ & $\dots$& $\dots$\\
$B_2$ & ppm & $-$3519(30) & $\dots$ & $\dots$& \\
$B_3$ & ppm & $-$215(32) & $\dots$ &  $\dots$& $\dots$\\
\hline 
\multicolumn{3}{l}{\quad Derived parameters} \\
$f_2$ & ppm &  670(120) & 1060(220) & $\dots$ & $\dots$ \\
$q$ & \dots & 0.116(12) & $\dots$ & $\dots$&  $\dots$  \\
$M_2$ & $M_{\rm Jup}$ & 102(12) & $\dots$ &  $\dots$& $\dots$ \\
$R_2$ & R$_{\rm Jup}$& 3.01(7) & 2.50(16)  &  2.76(16) & 1.84(17) \\
Irradiance &  W/m$^2$ & 6.41(80)$\times$10$^{6}$ & 3.12(54)$\times$10$^{6}$  &  2.94(48)$\times$10$^{4}$ & 5.6(17)$\times$10$^{4}$
\label{tab:StellarParamsNotPlanets}
\enddata
\tablenotetext{\dagger}{These stellar radius values are from \citet{morton2016} and are listed for comparison. \texttt{Isochrones} values were derived using Gaia EDR3 parallaxes \citep{gaiaEDR3} and publicly available photometric values.}
\tablenotetext{}{The values presented in the bracket is the error in the last 1--3 digits of the reported values. The reported error may straddle across the decimal point.}
\end{deluxetable*}

\begin{figure}[t]
    \centering
    \includegraphics[width=.45\textwidth]{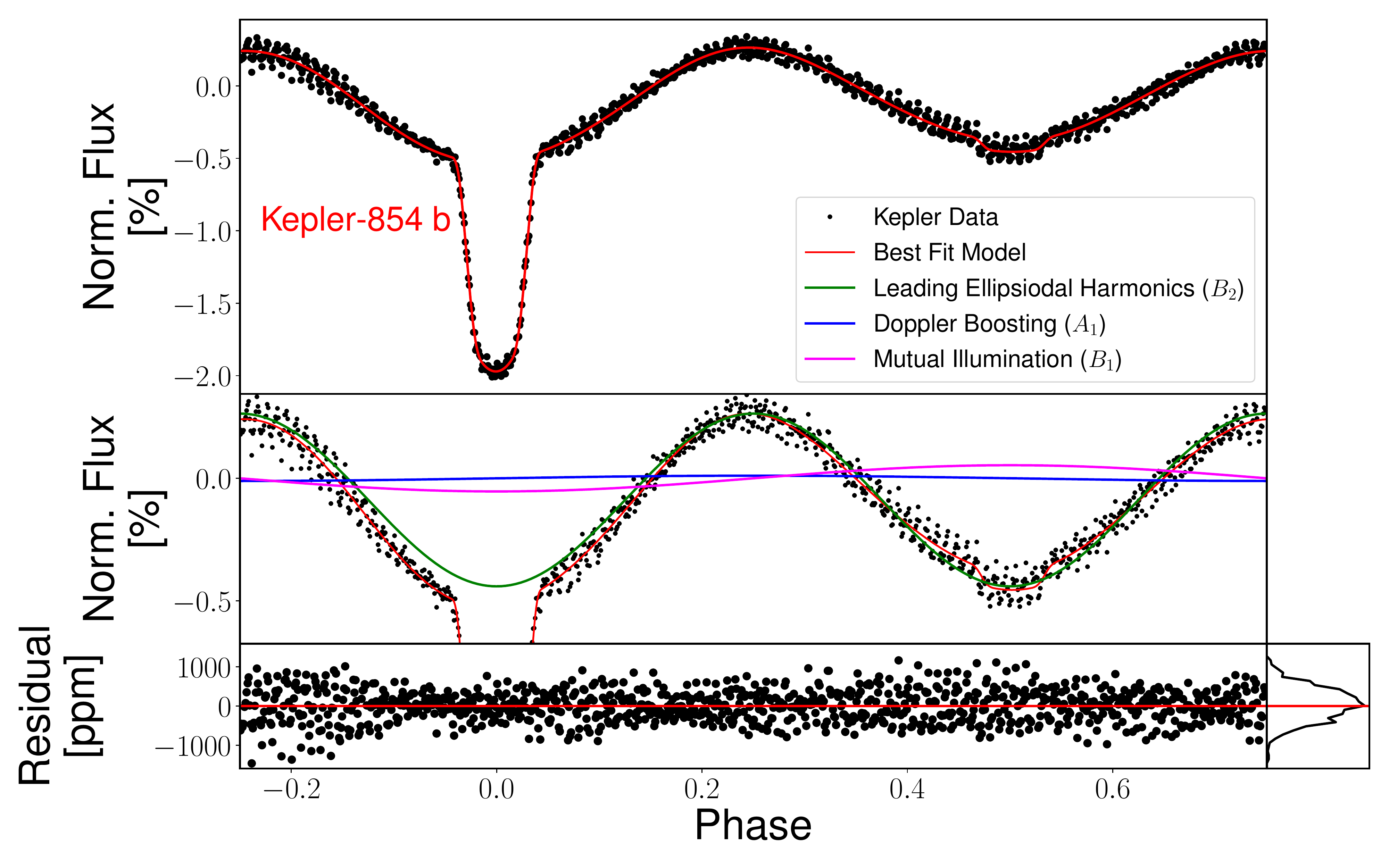}
    \includegraphics[width=.45\textwidth]{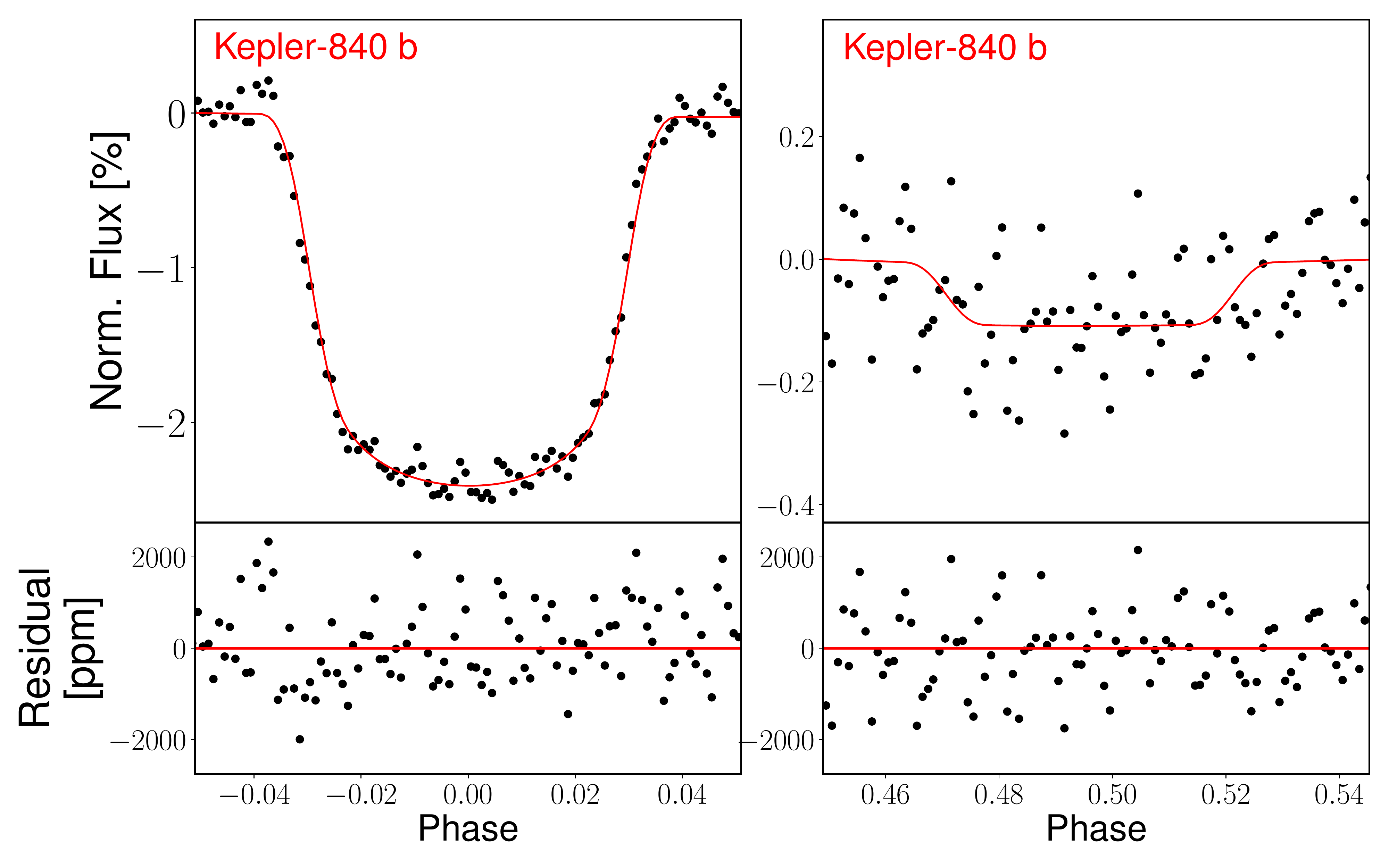}
    \includegraphics[width=.45\textwidth]{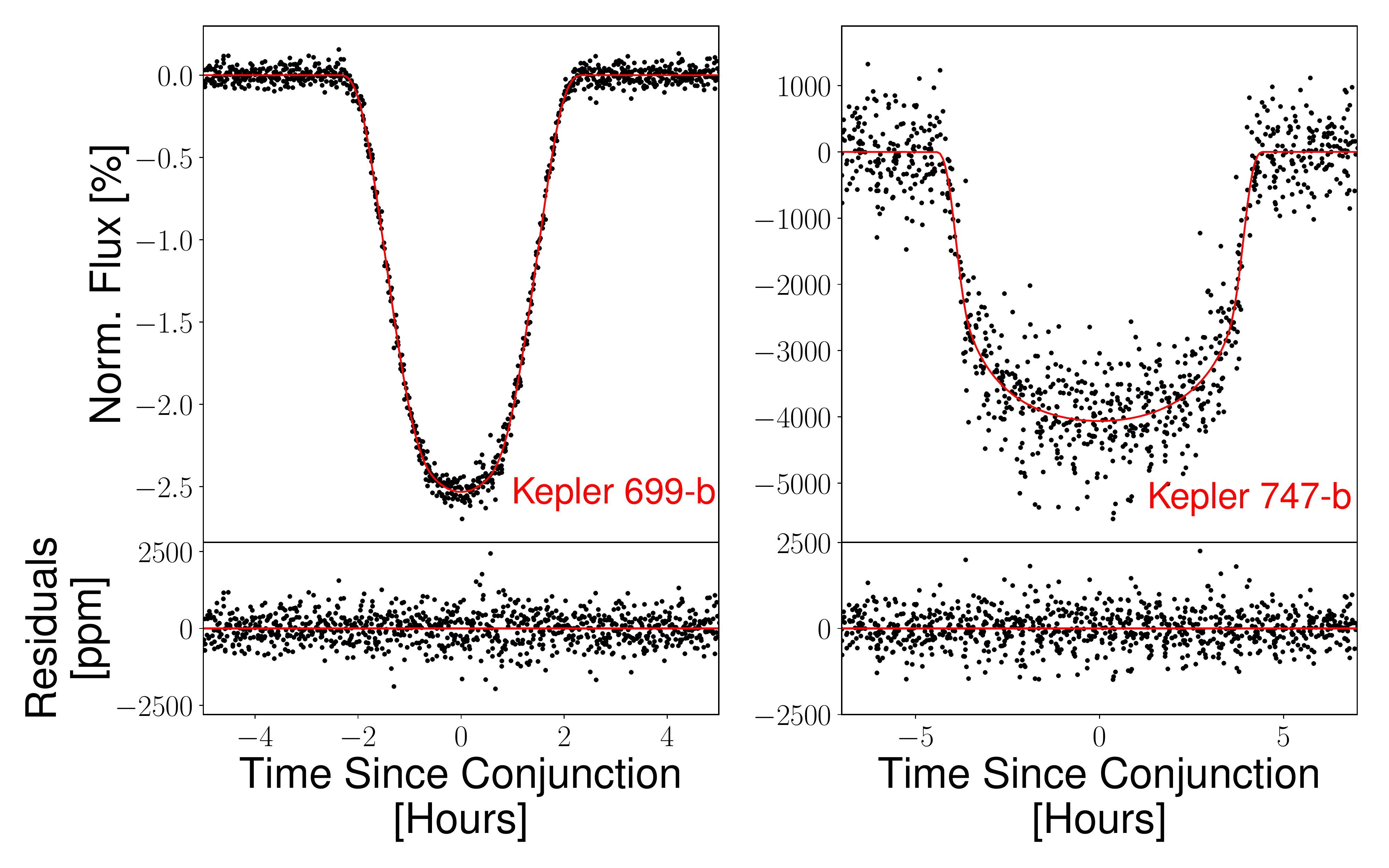}
    \caption{Top: the phase-curve signal extracted from the light curve of Kepler-854~b. The first panel shows the phase-folded, binned, and systematics-corrected light curve, with the best-fit model plotted in red. The second panel is a zoomed-in view of the phase modulation highlighting different components. The bottom panel is the residuals from the best-fit model, along with a histogram of the residuals. The large ellipsoidal variation amplitude ($3519 \pm 30$ ppm) indicates that the companion is not of planetary mass. Middle left: primary transit of Kepler-840~b from four years of Kepler data. Middle right: secondary eclipse of Kepler-840~b, with a measured depth of 1060$\pm$220 ppm. Bottom left: transit light curve of Kepler-699~b, along with the best-fit model and corresponding residuals. Bottom right: results from the transit light-curve fit for Kepler-747~b. }
    \label{fig:lightcurves}
\end{figure}

Below, we present the results of our full phase-curve analysis for Kepler-854~b, primary transit and secondary eclipse fit for Kepler-840~b, as well as the transit light-curve fit for Kepler-699~b and Kepler-747~b; the latter two systems have relatively long periods and hence do not show detectable phase-curve modulations. As for Kepler-840~b, we found that stellar activity strongly impacts the phase curve. The detailed data processing and fitting procedures used for the phase-curve and eclipse modeling analysis are described in Section~\ref{sec:fitting}.  The phase-folded light curves for these four systems are shown in Figure~\ref{fig:lightcurves}; the best-fit parameter values are listed in Table~\ref{tab:StellarParamsNotPlanets}.

\subsection{Kepler-854~b}

Kepler-854~b (KIC 7532973, TIC 275570329) was validated as a planetary candidate in  \citet{morton2016} (hereinafter M16). However, based on the range of estimated masses derived from the measured ellipsoidal effect in the Kepler photometry, we conclude that this companion is not a planet. In our analysis, we obtained statistically identical phase-curve amplitudes when splitting the Kepler light curve into two halves, indicating a consistent modulation across the time-series. We measured the amplitude of the leading term of the ellipsoidal variation (i.e., the second harmonic of the cosine) to be $B_2 = -3519 \pm 30$ ppm. Given the mass of the host star listed in the TIC ($0.84 \pm 0.04\,M_{\odot}$) we used Equation~\eqref{eqn:ellipsoidal_var_eqn}, a linear limb-darkening coefficient value of $0.581 \pm 0.040$, and a gravity-darkening coefficient value of $0.328 \pm 0.017$ from \citet{claret2011}, to derive a companion mass of $102 \pm 12\,M_{\rm Jup}$, i.e., likely a late-type dwarf star. This simplified calculation, which ignored the contribution of the companion's tidal distortion to the overall measured amplitude at the second harmonic, was supported by the low fitted relative brightness of the companion ($f_2 = 670 \pm 120$ ppm) and the small radius ratio ($R_2/R_1 = 0.1212 \pm 0.0019$). The radius ratio derived here is consistent with M16. 

The revised stellar radius bumps the host star's radius from $1.25^{+0.51}_{-0.17}\,R_{\odot}$ in M16 to $2.56 \pm 0.05\,R_{\odot}$ in our \texttt{isochrone} fitting, correspondingly raising the planet radius from $1.50^{+0.60}_{-0.20}\,R_{\rm Jup}$ to $3.01 \pm 0.07\,R_{\rm Jup}$ (see Table~\ref{tab:StellarParamsNotPlanets}). The bottom-right panel of Figure~\ref{fig:RadiusRatio} shows the planetary radius values of confirmed planets as reported in NASA Exoplanet Archive against the corresponding insolation flux. There is a strong correlation between planetary radius and insolation flux beyond $\sim$2$\times$10$^{5}$ W/m$^{2}$ \citep[e.g.,][]{demory2011}, but even among these inflated hot Jupiters, the physical radius of Kepler-854~b at $3.01 \pm 0.07\,R_{\rm Jup}$  stands out as being unusually large \citep{sestovic2018}, and therefore it is unlikely that Kepler-854~b is a planet. We note that our derived companion mass is low for an object of this size, which is expected to have a mass of $289 \pm 36\,M_{\rm Jup}$ according to the relationships in \citet{chen2017}. It is thus possible that the mass obtained from our \texttt{isochrones} analysis could be underestimated. If we instead use the stellar radius from the SED fitting, which is more accurately constrained than the stellar mass, and combine that with the density of $165 \pm 18$ kg m$^{-3}$ inferred from the transit fit, we obtain a host-star mass of $1.97 \pm 0.25\,M_{\odot}$, which would yield a companion mass of $239 \pm 54\,M_{\rm Jup}$ --- consistent with the expected mass of an object with the measured radius. In any case, we can rule out the planetary nature of Kepler-854~b at a high level of confidence.

\subsection{Kepler-840~b}

The strongest evidence against the planetary nature of Kepler-840~b (KIC 11517719, TIC 27847307)  comes from the revised physical radius of the planet. We report an updated host-star radius of $1.73 \pm 0.11\,R_{\odot}$ from  isochrone analysis (up from  $1.06^{+0.24}_{-0.12}\,R_{\odot}$ in M16), which correspondingly increases the measured planet radius from $1.52^{+0.35}_{-0.17}\,R_{\rm Jup}$ to $2.50 \pm 0.16\,R_{\rm Jup}$. The revised radius is too large for a substellar object (see bottom-right panel of Figure~\ref{fig:RadiusRatio}).

Additional evidence for the non-planetary nature comes from a large secondary eclipse depth of $1060 \pm 220$ ppm. Assuming reflection as the primary source of the secondary eclipse, this entails a geometric albedo of 1.37 $\pm$ 0.31.  This inferred geometric albedo is much higher than the values measured for hot Jupiters, which are close to $\sim$0.1 \citep[e.g.][]{esteves2015, angerhausen2015, wong2020Tess, wong2021}, or any solar system objects \citep{dyudina2016}. If we take the secondary eclipse to stem entirely from thermal emission, then the temperature of the secondary is $3455 \pm 150$ K, which is inconsistent with the equilibrium temperature of $2465 \pm 105$ K, when assuming zero Bond albedo and a recirculation efficiency ($\epsilon$) of 0 \citep{cowan2011}. Thus, this further supports the conclusion that the companion is a self-luminous object, i.e. a star, and not a planet. We attempted to measure the phase curve but found that it is strongly impacted by stellar activity. Specifically, the phase-curve fits from the two halves of the Kepler time series are not consistent with one another.

\subsection{Kepler-699~b}

Kepler-699~b (KIC 6061119, TIC 168811723) has an orbital period of 27.8 days, and at such a distance the phase-curve signal is not strong enough to be robustly detected in the Kepler light curve. We instead fit the transits only. With the revised parallax (Gaia EDR3), our \texttt{isochrone} analysis revealed the stellar radius to be $1.71 \pm 0.10$ $R_{\odot}$, up from $0.780^{+0.077}_{-0.067}$ $R_{\odot}$ reported in M16. This increase in the stellar radius yields a planet radius of $2.76 \pm 0.16\,R_{\rm Jup}$, unphysical for a planetary body, especially given that the companion is not highly irradiated and is therefore unlikely to be inflated to such an extent (see bottom-right panel of Figure~\ref{fig:RadiusRatio}).

\subsection{Kepler-747~b}

Kepler-747~b (KIC 5780460, TIC 121602473) is a long-period planet (35.6 days) that requires reexamination in light of the newly available parallax measurement of the host star. The updated stellar radius is $3.20 \pm 0.29$ $R_{\odot}$, revised from $0.780 \pm 0.042$ $R_{\odot}$ in M16, a factor of four increase. This leads to a companion radius of $1.84 \pm 0.17$ $R_{\rm Jup}$, an unlikely value given the irradiation level of the companion, as shown in the bottom-right panel of Figure~\ref{fig:RadiusRatio}. We note that the three planets in the vicinity of Kepler-747~b in the plot are Kepler-447~b \citep{kepler447}, NGTS-1~b \citep{bayliss2018}, and TOI-1130~c \citep{TOI1130c}, all of which have large radii uncertainties due to the grazing nature of their transits.  In the absence of a clear secondary eclipse, as in the case of Kepler-840~b, we cannot completely rule out Kepler-747~b as a planet based on its radius alone.

\section{Phase-curve Analysis of Kepler Eclipsing Binaries} 
\label{sec:dataanal}

In the previous section, we found that for Kepler-854~b, the complementary insights provided by the phase curve offer a powerful new perspective into the companions' properties. The phase curve of Kepler-854 shows a significant phase-curve modulation at the third harmonic, with $B_3 = -215 \pm 32$ ppm (Table~\ref{tab:StellarParamsNotPlanets}). Such high-order cosine harmonics have been detected in the phase curve of two planetary systems --- Kepler-13A and HAT-P-7  \citep{shporer2014, esteves2015} --- though at amplitudes that are at least an order of magnitude smaller. While it has generally been assumed that such higher harmonics are manifestations of the higher-order terms in the host star's ellipsoidal variation (see Section~\ref{sec:ellipsoidal}), the astrophysical nature of these signals has not been investigated empirically for a larger sample (for a numerical study, see \citealt{gomel2021}). 

Due to the lack of such systematic studies and the fact that only a few planetary systems have shown higher than expected amplitudes for the third cosine harmonics, the origins of such harmonics have come under debate. An earlier attempt to look at these higher harmonics among Kepler Objects of Interests by \citet{armstrong2015} suffered from both contamination from stellar activities and artifacts from the data analysis \citep{cowan2017}. This in turn inspired alternative explanations, such as consequences of eccentricity/inclination, pipeline artifacts, and effects of obliquity or weather, although tidal forces remain the most plausible explanation \citep{cowan2017, cowan2018}. The planetary phase curve picture is also further complicated by possible temporal weather patterns \citep{menou2009, armstrong2016, cho2021} or asymmetric cloud coverage \citep{demory2013, shporer2015}, which can lead to power leakage from lower stronger harmonics (often $B_1$ for planets) to higher order terms. Given the long observational baseline for Kepler, these variability should however average out. In this section, we supplement our previous analysis of misclassified planetary systems with an investigation of a sample of short-period Kepler eclipsing binaries, particularly those with high signal-to-noise phase-curve modulations. Analyzing these signals may shed new light on the origin of these harmonics and help contextualize similar high-harmonic phase-curve signals in other photometric observations.

\subsection{Target Selection}
\label{sec:targetselection}

We started with the Kepler eclipsing binary (EB) catalog, which contains 2,920 EBs\footnote{\url{http://keplerebs.villanova.edu}} \citep{prsa2011, conroy2014, kirk2016}. From that list, we selected EBs with orbital periods between 1.5 and 10 days, which are expected to be detached while expecting strong phase-curve signals. We also established a maximum Kepler magnitude of 13.0 to ensure adequate photometric precision. Next, we removed systems that do not show clear eclipses (i.e., contact binaries) and systems with eccentric orbits (i.e., where the secondary eclipse is not half an orbital period away from the primary transit). 

To ensure the presence of a coherent phase curve signal in the observed light curve, we divided the light curve into two halves (just as we did for the Kepler-854 and Kepler-840 light curves) and compared the binned phase curve from the two halves, removing any systems which showed observable discrepancies. This test helps us to assess the impact of stellar activity on the phase curve. The Kepler EB catalog was supplemented by other published studies of Kepler eclipsing binaries \citep{rappaport2015, faigler2015}. We note that the target selection criteria were relaxed somewhat for a few targets, such as KIC 4169521, a white dwarf system with a previously studied phase curve \citep{faigler2015}. Some of the targets show increased scatter in the residuals around the eclipses, which are potentially due to eclipse-timing variations, spot-crossing events, or imperfectly corrected dilution in some Kepler quarters \citep{stumpe2012}. These targets were also removed. 

\begin{figure}[t!]
\includegraphics[width=0.45\textwidth]{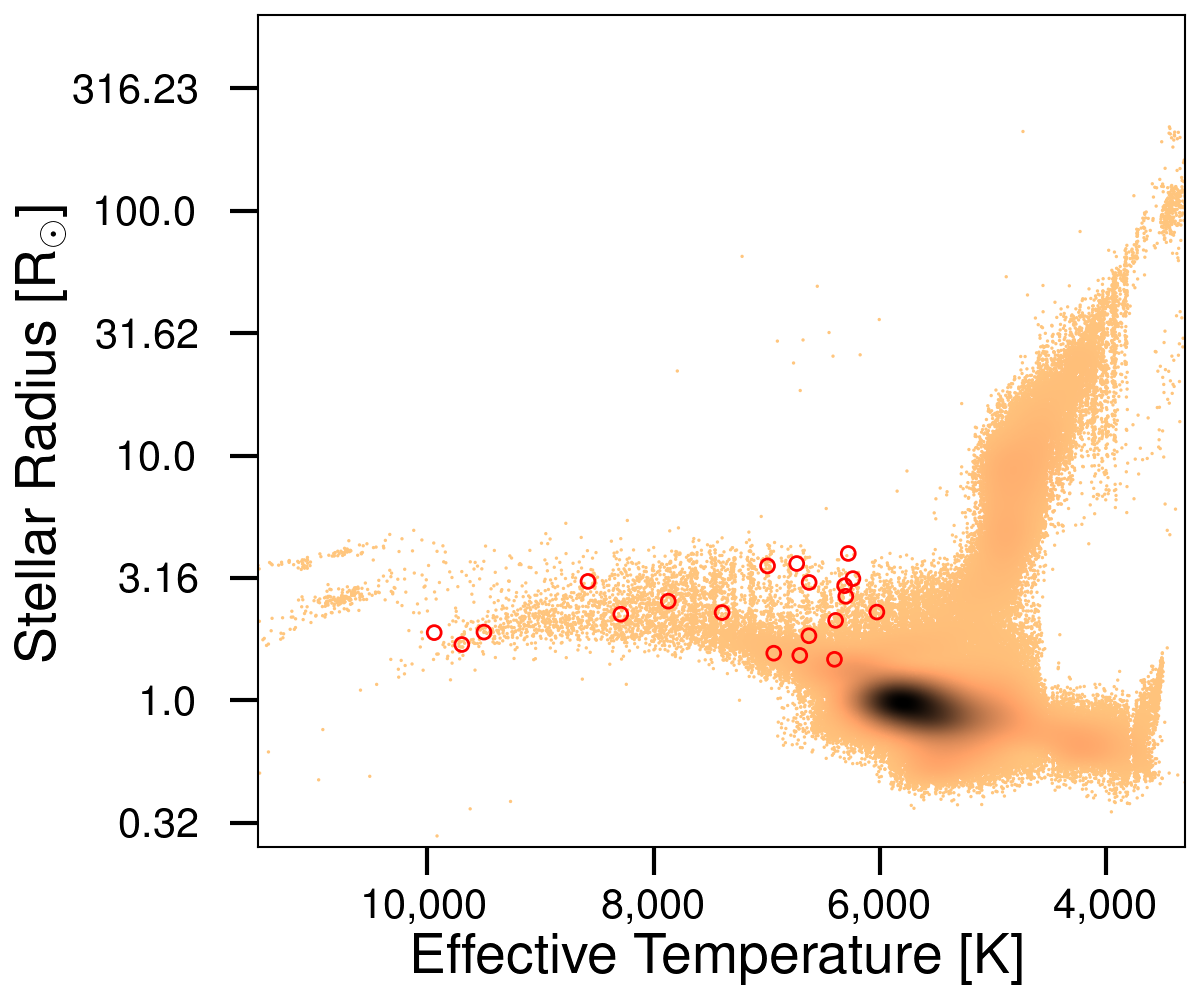}
\caption{\label{fig:TargetSelection} The distribution of our sample of 20 Kepler eclipsing binary systems (red circles) plotted on top of a color map showing the distribution of $\sim$200,000 targets from the Kepler mission, which primarily looked for planets around Sun-like stars. For the Kepler targets, the plotted parameters were taken from the KIC, whereas for our selected 20 EB systems, we used the more recent values from the TIC. }
\end{figure}

We performed a preliminary fitting analysis on around 60 EB systems, out of which 20 had both a radius ratio less than 50\% and a secondary flux contribution less than 10\%. As discussed in Section~\ref{sec:model}, when $R_2/R_1$ and $f_2$ are small, the contribution of the secondary companion to the overall phase-curve signal is negligible, allowing us to focus on the physical properties of the primary. These 20 systems we selected for our analysis are shown in Figure~\ref{fig:TargetSelection} in the context of the $\sim$200,000 Kepler targets. It is evident that our targets are biased toward hotter host stars. This bias could be introduced due to the likelihood of the cooler binary pairs developing co-rotating spots, which can destroy the long-term coherency of the phase-curve signal \citep{giles2017, rackham2019}.

\subsection{Model Fitting} 

\label{sec:fitting}
We downloaded all of the long cadence photometry available for each target from the Mikulski Archive for Space Telescopes (MAST) and used the latest processed light curves (DR 25; \citealt{jenkins2010, jenkins2017}). For our fitting analysis, we took the simple aperture photometry (SAP) light curve provided in the data files and discarded all data points with a nonzero quality flag. We corrected the light curves for long-term systematic trends by applying a cubic spline with knots tied every orbital period to the transit-masked photometry \citep{niraula2018, daylan2021}. To clean the photometry of flux ramps and any edge effects incurred by the spline detrending process, we removed the data affected by these spurious features that typically occurred near gaps longer than 6 hours. Finally, to further identify any remaining outliers or significant noise features, we phase-folded the light curve on the orbital period to build a template of a coherent periodic signal. We then used this template to construct the common-mode time series signal throughout the entire light curve and removed all data points that deviated by more than $3.5\sigma$. This step removed around 3\% of the time series, on average. 

When examining the Lomb--Scargle periodograms of the cubic spline models, we found that they include signals at the first harmonic of the orbital frequency at the level of $\sim$1\% of the total ellipsoidal amplitudes, on average. For targets with large photometric variations, such as KIC 8906039, this power leakage is comparable to or even exceeds the uncertainties on the measured amplitudes. A workaround was therefore developed to address this issue, which we describe later in this section. 

To characterize the system, we performed a simultaneous fit of the phase curve, primary transits, and secondary eclipses. The primary transits and secondary eclipses were fit using \texttt{ellc} \citep{maxted2016}, with a supersampling factor of 10 to account for Kepler's long-cadence integration time of 29.4 minutes. Our model accounts for the finite light-travel time across the orbit, which affects the measured secondary eclipse phase relative to the transit phase. We fit for the first five harmonics of both cosine and sine modulations, as described in Equation~\eqref{eqn:phasemodel}. As the input to our fitting pipeline, we used the median-binned, phase-folded time series, a process that expedites the fitting process relative to fitting the full unbinned, non-phase-folded photometry and naturally removes any significant non-periodic signals.

\begin{figure*}[ht]
\centering
  \begin{tabular}{lr}
    \includegraphics[width=.48\textwidth]{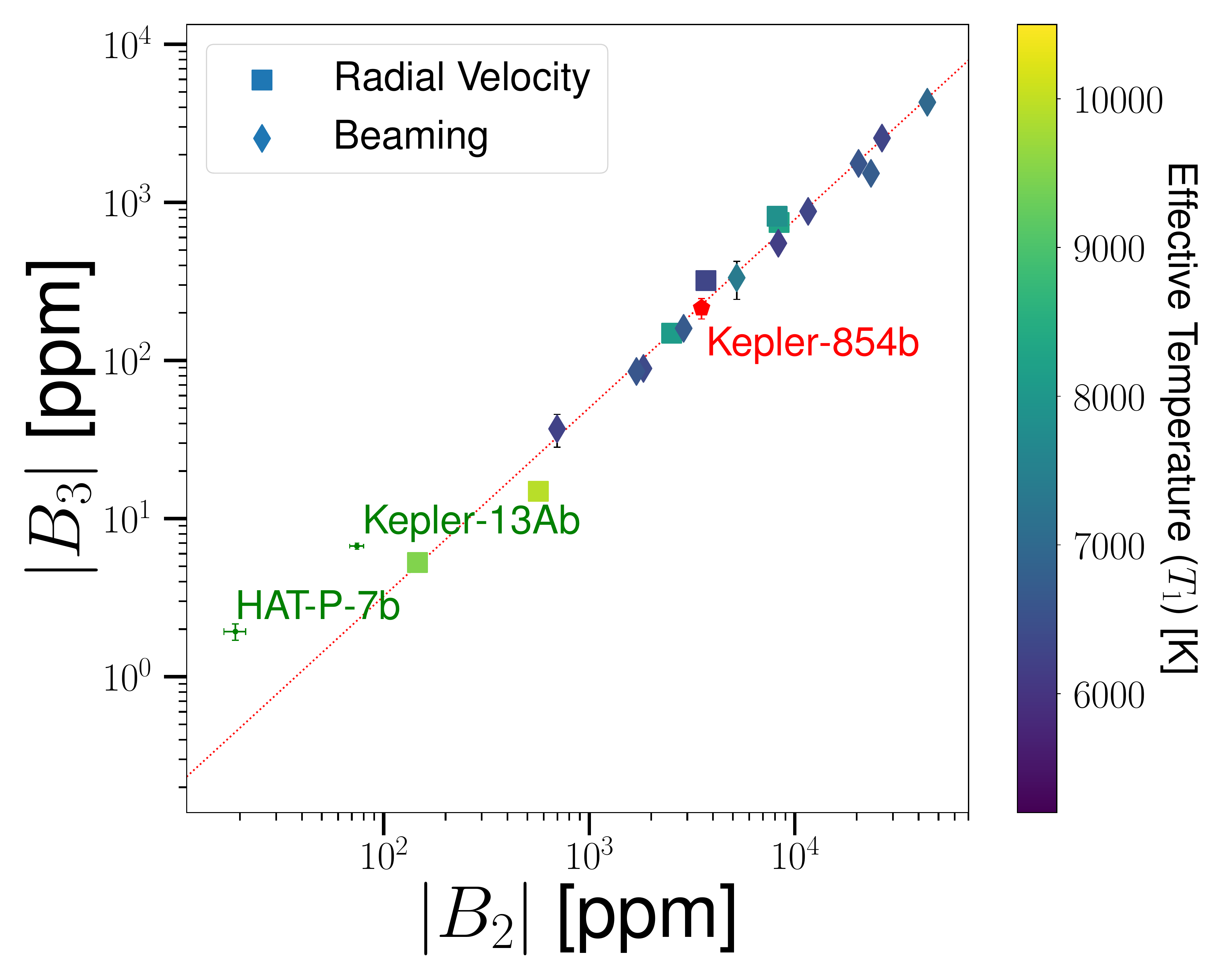} &
    \includegraphics[width=.48\textwidth]{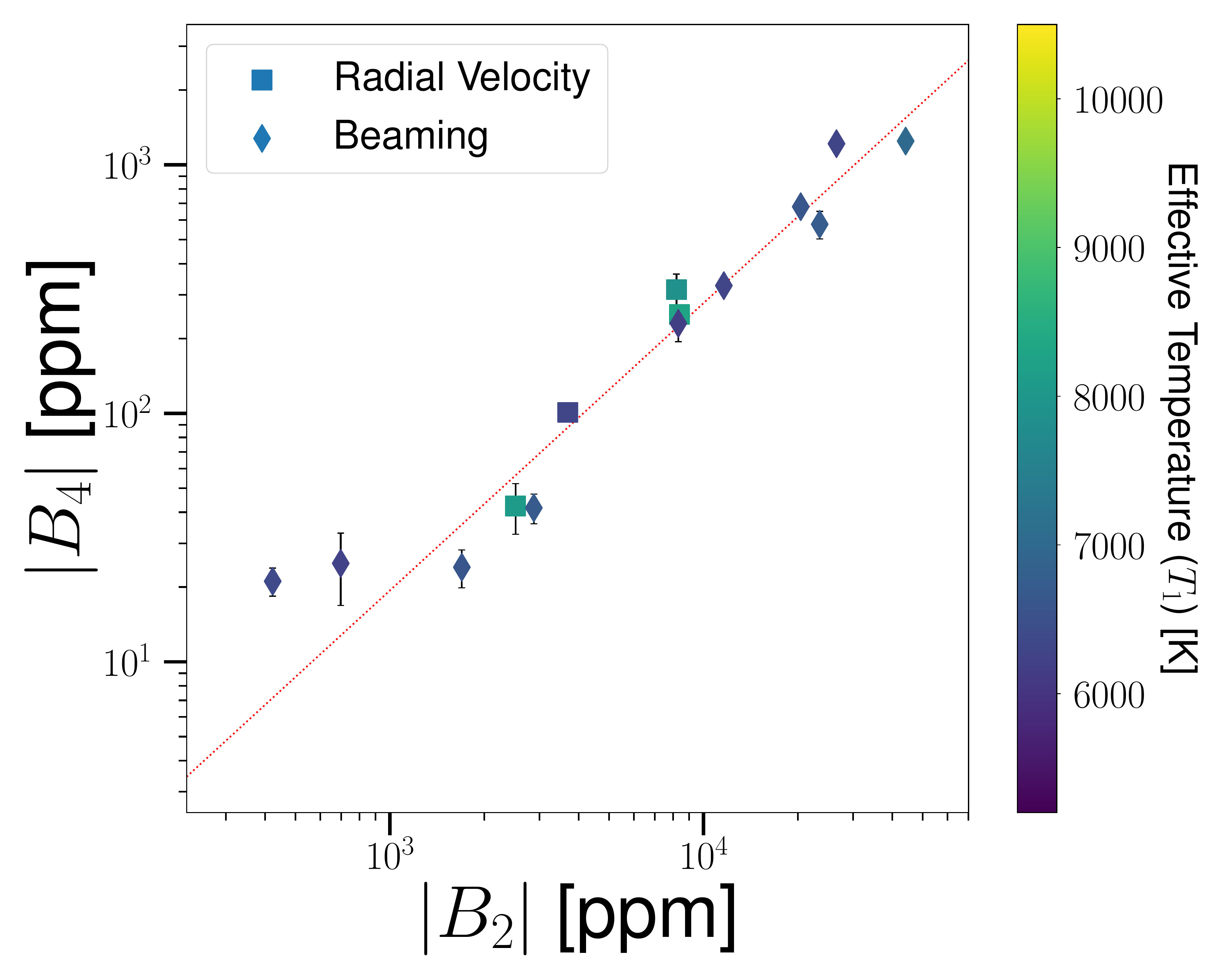} \\
  \end{tabular}
   \caption{\label{fig:HarmonicsRelationB2B3B4} Left: a comparison of the measured second- ($B_2$) and third-harmonic ($B_3$) cosine amplitudes among our sample of Kepler eclipsing binaries and planet-hosting systems, demonstrating a strong power-law relation between the two. The symbols indicate the method by which the system mass ratio $q$ was determined --- from the radial velocity signal or the simultaneously measured beaming amplitude $A_1$. The color scheme corresponds to the effective temperature of the primary star. The error bars are typically smaller than the markers. Right: same, but for the second- and fourth- harmonic cosine amplitudes. The two targets for which the signals are not statistically significant are not included.}
\end{figure*}

\begin{figure*}[ht]
\centering
  \begin{tabular}{lr}
    \includegraphics[width=.48\textwidth, trim={0 0.5cm 0 0.5cm},clip]{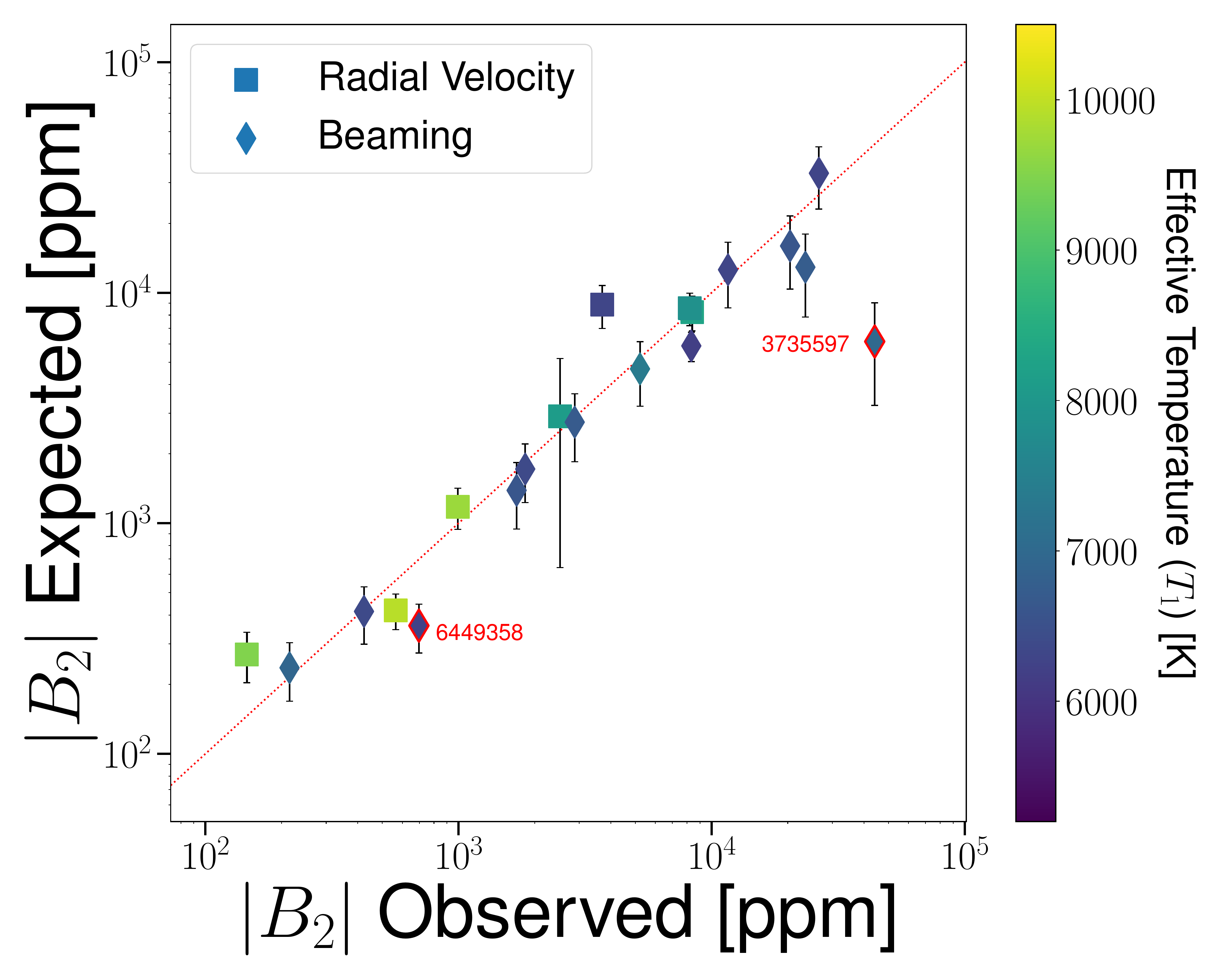}&
    \includegraphics[width=.48\textwidth, trim={0 0.5cm 0 0.5cm},clip]{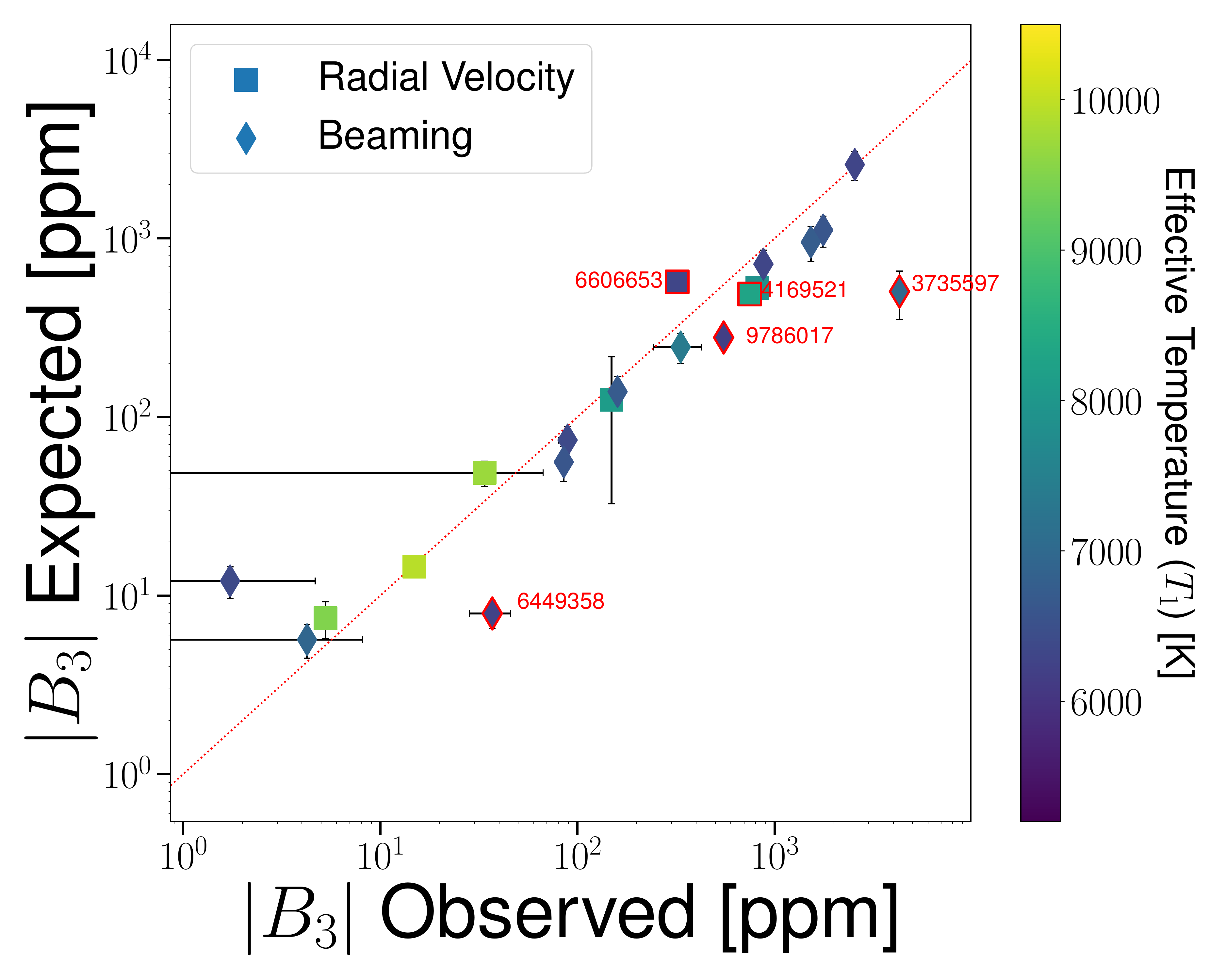}
  \end{tabular}
   \caption{\label{fig:TheoreticalB2B3} Left: a comparison of the measured and expected second-harmonic cosine phase-curve amplitudes ($B_2$), showing good agreement. The two systems (KIC 6449358 and KIC 3735597) that show a larger than $3\sigma$ discrepancy are marked with red outlines and labeled. Right: same, but for the third-harmonic cosine amplitudes ($B_3$). There are an additional three outliers marked in red.}
\end{figure*}

\begin{figure}
    \centering
    \includegraphics[width=.45\textwidth, trim= 0.25cm 0cm 0cm 0cm, clip]{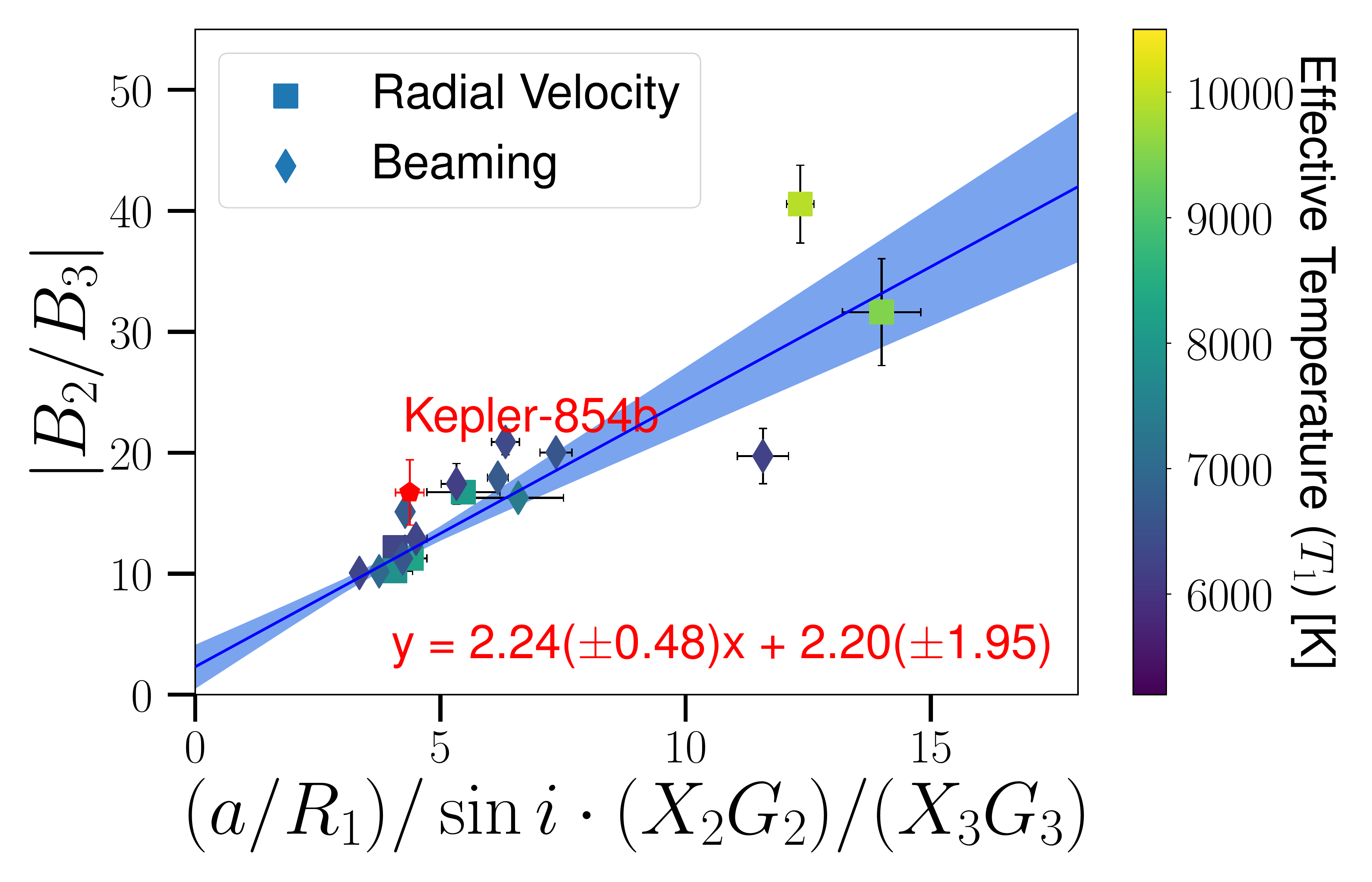}
    \caption{Observation of the expected linear relation between the ratio of second- and third-harmonic cosine amplitudes $|B_{2}/B_{3}|$ against a term containing the scaled semimajor axis, inclination, limb-darkening coefficients, and gravity-darkening coefficients (see Equation~\eqref{eqn:B2B3Ratio}). The best-fit line has a slope $2.24 \pm 0.48$, shown in blue, which is consistent with the theoretically expected value of 2.4. The shaded region shows the $1\sigma$ range of the fitted model. Three targets with statistically insignificant third harmonics are excluded from the fit.}
    \label{fig:RatioFigure}
    \vspace{+0.2cm}
\end{figure}

We used the Markov chain Monte Carlo (MCMC) ensemble sampler \texttt{emcee} \citep{emcee} with 22 parameters, including the radius ratio $R_2/R_1$, mid-transit time $T_0$, impact parameter $b$, scaled semimajor axis $a/R_1$, the luminosity ratio $S_2/S_1$ (which is directly related to the brightness ratio $f_2$), orbital period $P$, eccentricity parameters $e\sin\omega$ and $e\cos\omega$ (where $e$ is the orbital eccentricity and $\omega$ the argument of periastron), quadratic limb-darkening coefficients for both binary components (4 parameters total), and the phase-curve amplitudes (10 parameters). The host stars were assumed to be spherical; any deviation from sphericity due to ellipsoidal tidal distortion is expected to be at the level of a few percent at most, introducing negligible biases to our retrieved parameters. As for the limb darkening, we used the uninformative sampling technique and reparametrization developed by \citet{kipping2010}. We initialized the period and transit time at the values provided by the NASA Exoplanet Archive.  

To explore the parameter space, we ran the MCMC with 44 walkers (twice the number of parameters) for about 30,000 steps each. We ensured that the walkers were well-mixed. Once we confirmed that the fits were satisfactorily converged, we removed the burn-in phase. The burn-in phase was identified by visually examining the evolution of the log probability with the number of steps and determining the point when the walkers converged at the global log probability maximum. This typically occurred after about 15,000 steps, and we discarded the chain segments prior to this point before constructing the posterior distributions. 

In order to address the aforementioned contamination of the measured phase-curve signal from the spline detrending process, we tried fitting the uncorrected, non-phase-folded light curves, after masking the transits and eclipses. We used the priors on the transit ephemeris from the previous full phase-curve fits and then ran an additional set of MCMC fits for all 20 systems. The typical uncertainties on the phase-curve amplitudes from fitting the undetrended light curves were larger by a factor of two when compared to the detrended, phase-folded light-curve fits. However, the methods showed broad mutual consistency, except for a few cases such as KIC 8906039. In the Kepler EB phase-curve results presented in this paper, we choose the more conservative approach and report the best-fit amplitudes from the undetrended light-curve fits.

\subsection{Results}
\label{sec:res}


For our 20 selected systems, the phase-folded light curves with best-fit models overplotted are shown in Figures~\ref{fig:BestFit1} and \ref{fig:BestFit2}, and the fitted parameters are listed in Tables~\ref{tab:KepEB1} and~\ref{tab:KepEB2}. As reported in these tables, the first cosine harmonic amplitudes ($B_1$), which is dominated by the mutual illumination, are typically negative. However, some systems show positive $B_1$ values. These systems all have brightness ratios ($S_2/S_1>1$) and correspond to the EBs that have hot white dwarf secondaries. This behavior is expected, as the sign of the mutual illumination amplitude is driven primarily by the temperature ratio of two binary components. As for the higher-order cosine amplitudes, we find clear empirical power-law relations between the second ($B_2$) and third ($B_3$) cosine harmonics (left panel of Figure~\ref{fig:HarmonicsRelationB2B3B4}) as well as between the second and fourth ($B_4$) cosine harmonics (right panel of Figure~\ref{fig:HarmonicsRelationB2B3B4}). Such a positive correlation is expected from Equation~\eqref{eqn:ellipsoidal_var_eqn}, as all the cosine harmonics scale proportionately with the mass ratio, and inversely with the scaled distance.  Our data set encompasses measured phase-curve amplitudes that span two orders of magnitude, often recommended for claiming robust power law. The scatter can be attributed to the range of inclinations, gravity- and limb-darkening coefficients, as well as the differing contributions from the secondary companions. The observed power law suggests a common provenance for the signals in these systems and strongly implies that the higher cosine harmonics (both $B_3$ and $B_4$), just as the second cosine harmonic ($B_2$), originate from the mutual tidal interaction between the binary components. 

In order to find the best-fit power law, we used the standard distance regression routine in \texttt{scipy}. To estimate the uncertainties in the fit parameters, we performed a fit to half of the data points at a time, carried out the regression on all the possible subsets, and calculated the median and standard deviation from the resultant distribution of fitted values. In the left panel of Figure~\ref{fig:HarmonicsRelationB2B3B4}, we include Kepler-854, which is positioned close to the fitted relation, further highlighting the non-planetary nature of the phase curve, as planets primarily occur at the bottom-left corner of the plot. Notably, the planetary systems located at the bottom-left corner in Figure~\ref{fig:HarmonicsRelationB2B3B4} deviate from the expected empirical law, which could be due to additional unconsidered phenomena, including those mentioned previously at the beginning of this section.

Figure~\ref{fig:TheoreticalB2B3} presents a comparison between the measured and the corresponding expected values given the system parameters for the second (left panel) and third harmonic (right panel), assuming they originate from the ellipsoidal variation as described in Section~\ref{sec:ellipsoidal}. For each system, the mass ratio $q$ was estimated based on the published radial velocity orbits (if available) or by deriving it from the measured beaming amplitude (see Section~\ref{sec:beaming} and also Equation~1 in \citealt{shporer2010}). The ellipsoidal and beaming coefficients were derived using the stellar parameters from the TIC \citep{stassun2019}. Both panels in Figure~\ref{fig:TheoreticalB2B3} show good agreement between the measured and expected values, with consistency at a level close to or better than 1$\sigma$ for most systems. A few systems do show significant discrepancies ($\ge 3\sigma$) between the measured and expected values of $B_2$ in Figure~\ref{fig:TheoreticalB2B3}. These include KIC 3735597 and 6449358, where the observed amplitude is larger than the expected value. Within the precision of our data, we do not see more significant discrepancies for the hotter stars, as was predicted by \citet{Pfahl2008}. For all of these discrepant systems, the mass ratio was derived from the measured beaming amplitude ($A_1$). Such deviations have been reported for systems such as KIC 4169521 \citep{faigler2015} and KIC 9164561 \citep{rappaport2015}, which exhibit statistically significant differences between the mass based on the beaming amplitude and the mass derived from the radial velocity signal. This is perhaps due to the presence of a phase shift in the mutual illumination signal \citep{rappaport2015}, which would manifest as a systematic bias in the measured $A_1$ amplitude and translate into an incorrect beaming-derived mass estimate. For KIC 3735597 and 6449358, a systematic underestimation of the mass from beaming is a plausible explanation, as they simultaneously have larger than expected $B_3$ measurements. Additional targets such as KIC 4169521, 6606653, and 9786017 also show disagreement between the observed and measured $B_3$ amplitudes. It appears that while the classical estimation provides good order-of-magnitude predictions across the board, additional astrophysical phenomena may be contributing nonnegligibly towards the second- and third-harmonic amplitudes in some individual systems.

Given that obtaining a precise and accurate mass ratio estimate is difficult, we explored alternative pathways to further test the veracity of the classical formulation of ellipsoidal variation. As outlined in Section~\ref{sec:ellipsoidal}, we compared the expected ratio between $B_2$ and $B_3$, given in Equation~\eqref{eqn:B2B3Ratio}, to our measured value. The advantage of using $|B_2/B_3|$ is that it is approximately independent of the mass ratio for systems with relatively small and faint secondaries. Meanwhile, $a/R_1$ and $\sin i$ are fitted parameters from our light-curve analysis. In Figure~\ref{fig:RatioFigure}, we plot $|B_2/B_3|$ vs. $a/R_1/\sin i \cdot (X_2G_2)/(X_3G_3)$, where the expected relation should be linear with a slope of 2.4, following Equation~\eqref{eqn:B2B3Ratio}. We fitted a linear relation that yielded a slope of 2.24$\pm$0.48 and an intercept that is consistent with zero. Despite a large scatter, this value is statistically consistent with 2.4, the value expected from the classical formalism. We did not fix the y-intercept to zero, as it leads to an underestimation of the slope uncertainty.

\begin{deluxetable*}{rccccccc}
\tablecaption{\label{tab:KepEB1}Stellar Parameters for Selected Kepler Eclipsing Binaries}
 \tablehead{
 \colhead{KIC} & \colhead{Period (days)} & \colhead{$q$} & \colhead{$R_2/R_1$} & \colhead{$S_2/S_1$}
 & \colhead{$f_2$}
 }
 \startdata
2851474 & 2.768271874(11) & 0.0894(54)$^{a}$ & 0.01139(54) & 1.74(17) & 0.0002261(60)\\ 
3735597 & 1.96637378(11) & $\dots$ & 0.29830(93) & 0.4890(39) & 0.04350(22)\\ 
3836439 & 1.54039687573(48) & $\dots$ & 0.436(44) & 0.1727(39) & 0.0330(74)\\ 
4169521 & 1.172550761(37) & 0.1064(74)$^{a}$ & 0.0411(12) & 3.16(18) & 0.005345(18)\\ 
6449358 & 5.776793539(30) & $\dots$ & 0.32667(51) & 0.10050(36) & 0.010723(22)\\ 
6606653 & 2.698027122(36) & 0.230(20)$^{b}$ & 0.03923(33) & 8.51(14) & 0.013089(21)\\ 
6863229 & 1.9949176256(22) & $\dots$ & 0.36860(38) & 0.6606(15) & 0.089778(85)\\ 
6889235 & 5.1886710(18) & 0.1040(40)$^{c}$ & 0.02109(22) & 2.627(56) & 0.0011679(35)\\ 
7097571 & 2.213967147(26) & $\dots$ & 0.31804(63) & 0.2742(11) & 0.027737(49)\\ 
8112013 & 1.79051928652(63) & $\dots$ & 0.23686(43) & 0.04015(19) & 0.0022518(78)\\ 
8801343 & 2.73986948(11) & $\dots$ & 0.1370(10) & 0.0367(11) & 0.000688(19)\\ 
8878681 & 2.49633542(11) & $\dots$ & 0.1726(35) & 0.0282(13) & 0.000840(55)\\ 
8906039 & 2.34748845(11) & $\dots$ & 0.17356(31) & 0.2947(24) & 0.008876(59)\\ 
9071386 & 4.69328598(12) & $\dots$ & 0.11290(38) & 0.02189(78) & 0.0002789(98)\\ 
9164561 & 1.26704110(35) & 0.0970(40)$^{d}$ & 0.1133(22) & 2.298(90) & 0.02952(16)\\ 
9786017 & 4.4979773(62) & $\dots$ & 0.2419(14) & 0.931(12) & 0.05444(21)\\ 
10657664 & 3.273691223(62) & 0.106(12)$^{e}$ & 0.08128(11) & 2.8661(78) & 0.0189347(52)\\ 
10727668 & 2.305901907(37) & 0.150(10)$^{d}$ & 0.08465(69) & 3.096(50) & 0.022179(61)\\ 
11826400 & 5.889366939(20) & $\dots$ & 0.22497(35) & 0.02466(19) & 0.0012478(89)\\ 
12257908 & 2.61594468(55) & $\dots$ & 0.27153(81) & 0.4866(33) & 0.03590(14)\\ 
\enddata
\tablenotetext{}{References: a) \citet{faigler2015}  b) \citet{breton2012}  c) \citet{rowe2010}   d) \citet{rappaport2015} e)  \citet{WongKOI964_2020} }
\end{deluxetable*}

\begin{deluxetable*}{rccccccccccc}[!ht]
\tablecaption{\label{tab:KepEB2}Fitted Phase-curve Parameters for Selected Kepler Eclipsing Binaries}
\tablehead{
\colhead{KIC} & \colhead{$A_{1,{\rm Th}}$ (ppm)}&  \colhead{$A_1$ (ppm)}&  \colhead{$B_1$ (ppm)}& \colhead{$B_{2,{\rm Th}}$ (ppm)}& \colhead{$B_2$ (ppm)}& \colhead{$B_{3,{\rm Th}}$ (ppm)}& \colhead{$B_3$ (ppm)}&   \colhead{$B_4$ (ppm)}   
}
\startdata
2851474 & $156(21)$ & $128.2(11)$ & $-38.1(19)$ & $-2900(2300)$ & $-2260.9(19)$ & $-125(93)$ & $-135.0(19)$ & $38.6(17)$ \\ 
3735597 &  $\dots$  & $225(20)$ & $-3791(86)$ & $-6100(2300)$ & $-44017(60)$ & $-490(120)$ & $-4318(71)$ & $1273(45)$ \\ 
3836439 &  $\dots$  & $921.8(54)$ & $-2382.0(74)$ & $-5100(1500)$ & $-5041.3(60)$ & $-269(48)$ & $-309.8(65)$ & $67.2(66)$ \\ 
4169521 & $191(16)$ & $82.8(29)$ & $817.1(46)$ & $-8200(1400)$ & $-8411.7(39)$ & $-488(66)$ & $-747.3(39)$ & $225.0(39)$ \\ 
6449358 &  $\dots$  & $305.7(36)$ & $-180.2(45)$ & $-335(80)$ & $-740.2(42)$ & $-7.4(13)$ & $-38.0(42)$ & $0.7(41)$ \\ 
6606653 & $256(26)$ & $229.1(36)$ & $106.3(67)$ & $-8900(1900)$ & $-3713.5(53)$ & $-571(73)$ & $-305.8(60)$ & $98.7(50)$ \\ 
6863229 &  $\dots$  & $880.1(83)$ & $-2411(26)$ & $-14700(5100)$ & $-20299(19)$ & $-1020(200)$ & $-1801(22)$ & $640(14)$ \\ 
6889235 & $111.8(94)$ & $99.12(68)$ & $10.50(67)$ & $-270(67)$ & $-143.06(66)$ & $-7.5(17)$ & $-4.53(63)$ & $0.53(65)$ \\ 
7097571 &  $\dots$  & $1017.0(64)$ & $-1893(13)$ & $-13400(4300)$ & $-11589(11)$ & $-770(150)$ & $-899(12)$ & $309.7(92)$ \\ 
8112013 &  $\dots$  & $534.8(12)$ & $-497.9(19)$ & $-2770(900)$ & $-2874.1(17)$ & $-140(29)$ & $-160.2(19)$ & $38.6(16)$ \\ 
8801343 &  $\dots$  & $323.1(36)$ & $-181.2(47)$ & $-1700(490)$ & $-1857.9(45)$ & $-73(14)$ & $-89.1(44)$ & $18.6(43)$ \\ 
8878681 &  $\dots$  & $404.0(23)$ & $-251.4(19)$ & $-1350(430)$ & $-1698.0(25)$ & $-54(12)$ & $-85.0(27)$ & $30.6(24)$ \\ 
8906039 &  $\dots$  & $734.2(77)$ & $-2009(19)$ & $-26500(7700)$ & $-21976(16)$ & $-2080(370)$ & $-2184(17)$ & $1027(13)$ \\ 
9071386 &  $\dots$  & $196.4(21)$ & $-43.9(22)$ & $-390(110)$ & $-424.1(23)$ & $-11.4(23)$ & $-7.4(24)$ & $9.3(23)$ \\ 
9164561 & $151.6(67)$ & $-39(19)$ & $2344(40)$ & $-8600(1400)$ & $-8180(43)$ & $-529(44)$ & $-802(38)$ & $330(29)$ \\ 
9786017 &  $\dots$  & $191(34)$ & $-856(46)$ & $-1880(530)$ & $-8496(43)$ & $-82(16)$ & $-491(43)$ & $266(48)$ \\ 
10657664 & $106(15)$ & $96.68(92)$ & $264.6(11)$ & $-419(74)$ & $-572.0(10)$ & $-14.5(17)$ & $-14.2(11)$ & $2.38(72)$ \\ 
10727668 & $137(22)$ & $151(12)$ & $449(14)$ & $-1180(240)$ & $-992(14)$ & $-48.7(78)$ & $-46(14)$ & $6(14)$ \\ 
11826400 &  $\dots$  & $269.9(16)$ & $-57.6(18)$ & $-182(52)$ & $-227.0(18)$ & $-4.37(93)$ & $-3.9(18)$ & $5.2(18)$ \\ 
12257908 &  $\dots$  & $614(18)$ & $-1020(48)$ & $-10800(4200)$ & $-23611(36)$ & $-800(170)$ & $-1562(40)$ & $480(28)$ \\ 
\enddata
\tablenotetext{}{$A_1$, $B_1$, $B_2$, $B_3$, and $B_4$ are measured amplitudes from the light-curve fits, whereas the corresponding columns with the subscript `Th' are the theoretical predictions.}
\end{deluxetable*}

\section{Summary and Conclusions}
\label{sec:conclusion}
In this paper, we showed that three statistically validated transiting planets, Kepler-854~b, Kepler-840~b, and Kepler-699~b cannot be planets but are instead small stars. We used updated host-star parameters from the TIC and stellar SED fitting to show that the measured transit depth indicates a companion radius that is too large for a planet. For Kepler-854~b, we also used the Kepler phase curve to show that the transiting object's mass is inconsistent with its being a planet. In addition, we demonstrated that a fourth statistically validated planet, Kepler-747~b, has a somewhat larger radius than other confirmed gas-giant planets with the same stellar irradiation level, suggesting that it too is not a planet, but a small star.

This work demonstrated the use of phase curves as a validation tool for Kepler-854~b, as the phase curve amplitude is sensitive to the transiting object's mass (more directly, the system mass ratio). Meanwhile, the radius is broadly unchanged for objects from about 1~$M_{\mathrm{Jup}}$ ($\sim$0.001~$M_{\sun}$) to about 100~$M_{\mathrm{Jup}}$ ($\sim$0.1~$M_{\sun}$) \citep[e.g.,][]{chen2017}, making statistical validation methods that primarily rely on the companion radius susceptible to false positive scenarios. Fortunately, orbital phase-curve modulations can be detected automatically for a large number of objects using the techniques described here and elsewhere \citep[e.g.,][]{shporer2011, faigler2011, niraula2018, wong2020Tess}. 

Our work also shows the critical role that accurate host-star parameters play in validating transiting planet candidates \citep[see also][]{shporer2017vespa}. Here, we have benefited from measurements carried out by the ESA Gaia Mission \citep{GaiaDR2}, which were not available during the Kepler mission and have since dramatically improved the quality of stellar parameters for planet-hosting stars and, in turn, decreased the number of false positive candidates. 

Following the detection of a third-harmonic signal in the phase curve of Kepler-854, we used a sample of Kepler EBs to empirically investigate the nature of such higher-order signals. We found that the amplitude of the measured signal from Kepler-854 is consistent with the behavior of the Kepler EB sample. Moreover, the higher-order harmonic amplitudes measured from the EB sample show excellent agreement with the expected values from theoretical modeling of the tidal distortion of the primary stars. While perhaps somewhat expected, to the best of our knowledge, such an empirical study has not been done before, and our results provide us with a better understanding of not only the Kepler-854 phase curve, but also the contribution of ellipsoidal variation to visible-light phase curves in general.

With the increasing number of transiting planet candidates from the NASA TESS Mission \citep{guerrero2021} and the already limited amount of observational resources, statistical validation of candidates is becoming more and more essential. Some of the lessons learned in this work will help improve statistical validation techniques, which will, in turn, support more accurate studies of planet demographics.

\acknowledgments
This paper includes data collected by the Kepler mission and obtained from the MAST data archive at the Space Telescope Science Institute (STScI). Funding for the Kepler mission is provided by the NASA Science Mission Directorate. STScI is operated by the Association of Universities for Research in Astronomy, Inc., under NASA contract NAS 5–26555.
This research made use of the NASA Exoplanet Archive, which is operated by the California Institute of Technology, under contract with the National Aeronautics and Space Administration under the Exoplanet Exploration Program. 
\facilities{Kepler} 
\software{
astropy \citep{astropy}, 
emcee \citep{emcee}, isochrones \citep{morton2015}, 
ellc \citep{maxted2016}, \texttt{matplotlib} \citep{matplotlib},
\texttt{scipy} \citep{scipy}.} 
\pagebreak
\appendix
\label{sec:appendix}

\begin{figure*}[htb]
\centering
  \begin{tabular}{@{}ccc@{}}
    \includegraphics[width=.315\textwidth]{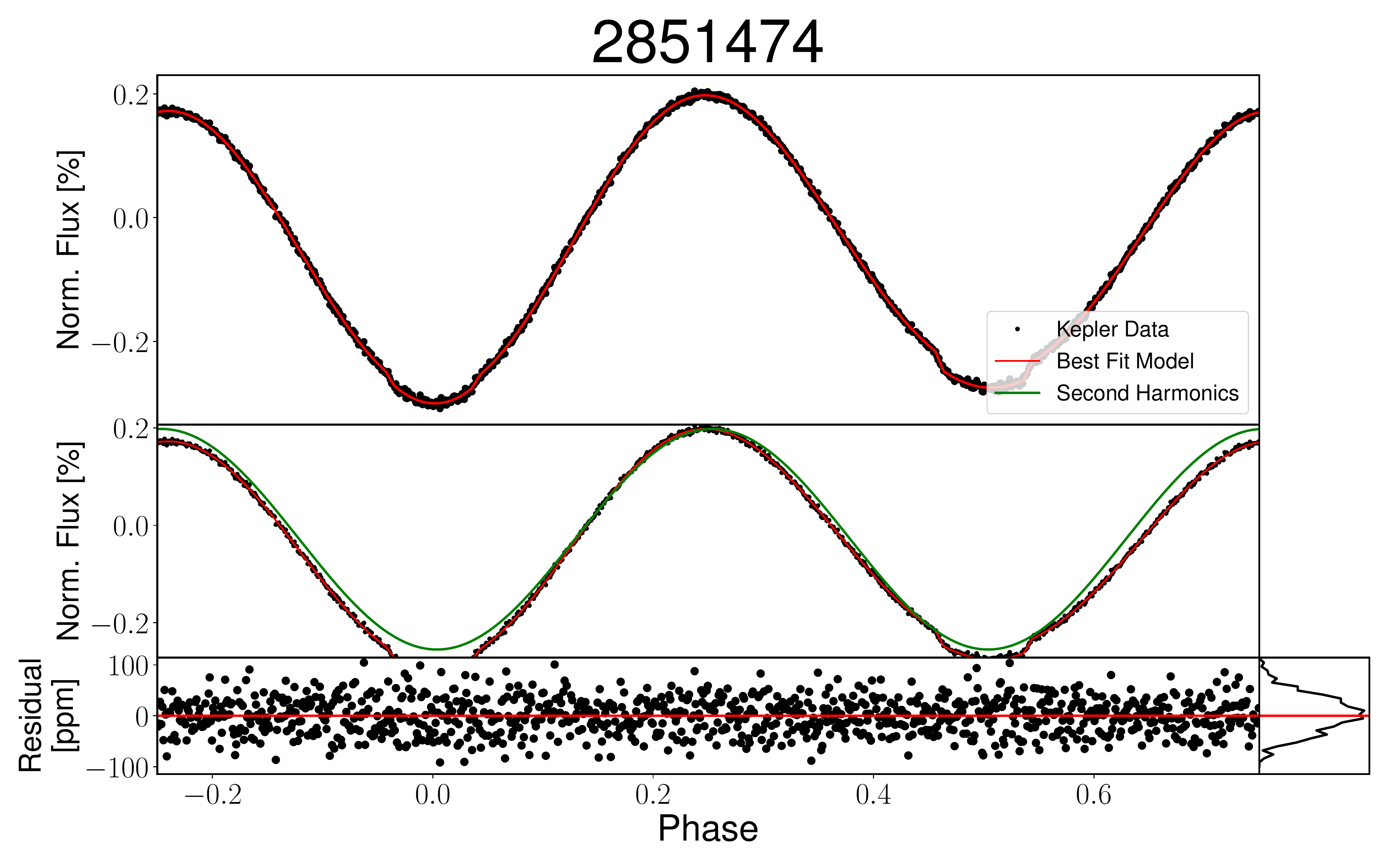}& 
    \includegraphics[width=.315\textwidth]{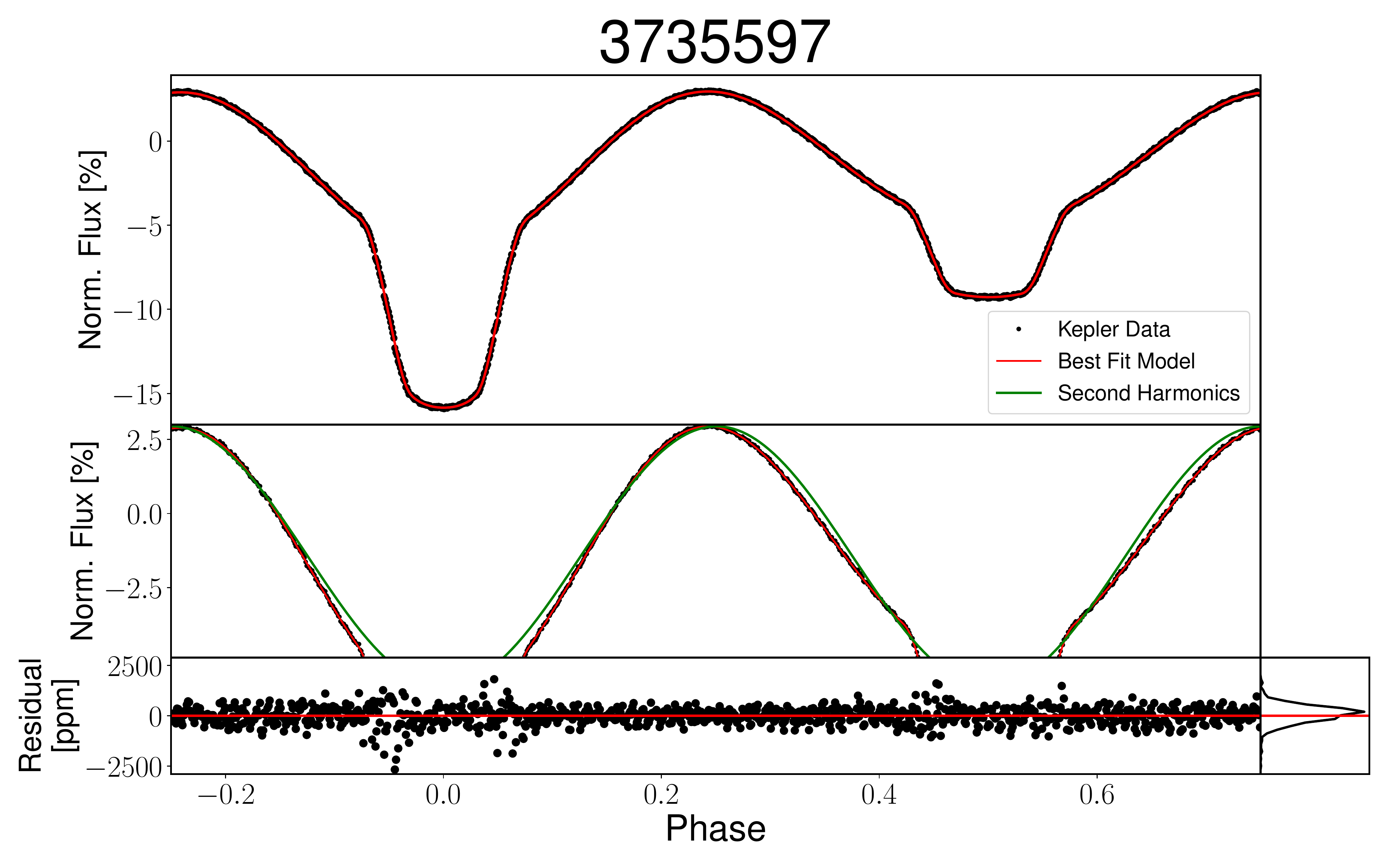} &
    \includegraphics[width=.315\textwidth]{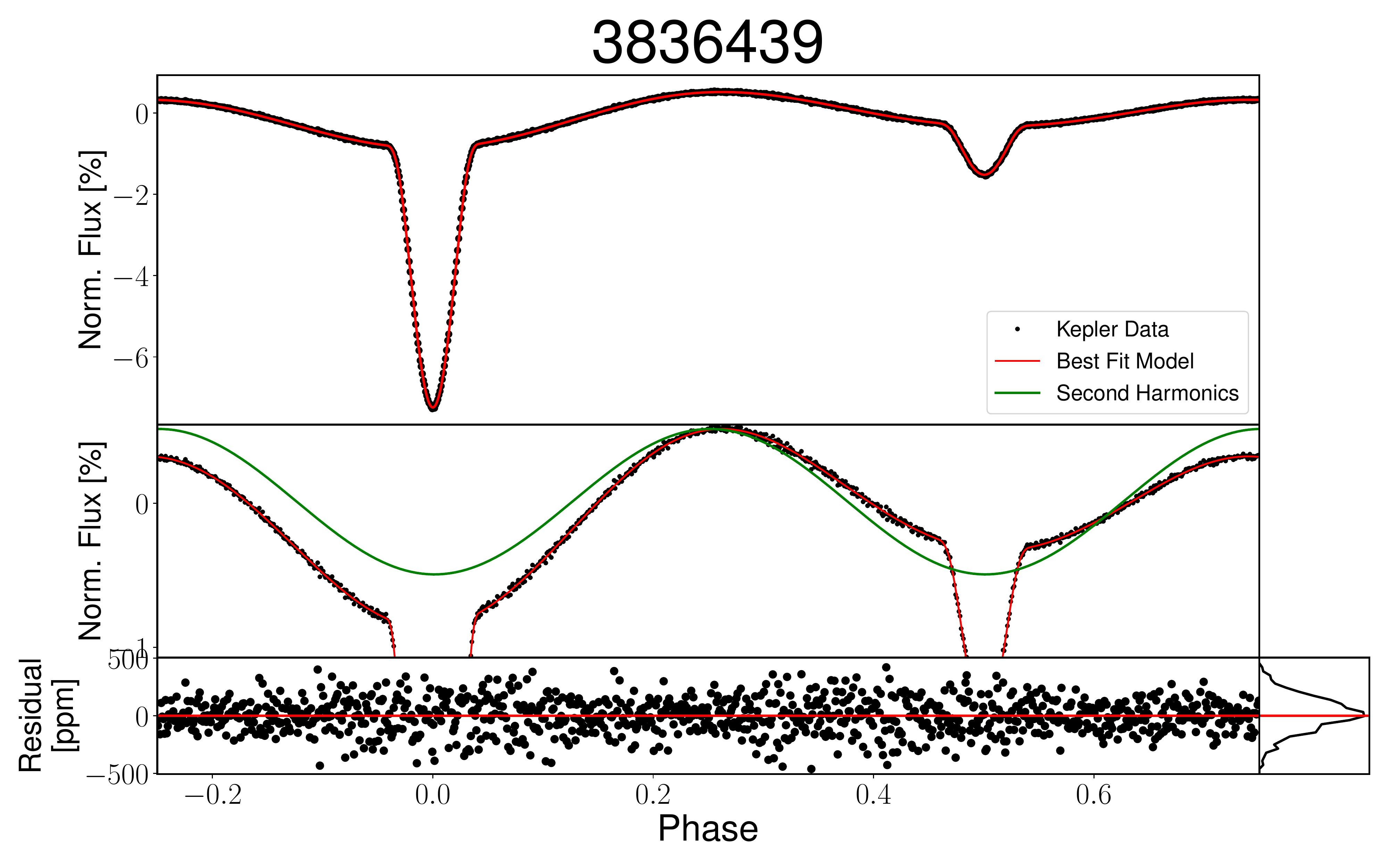}\\
    \includegraphics[width=.315\textwidth]{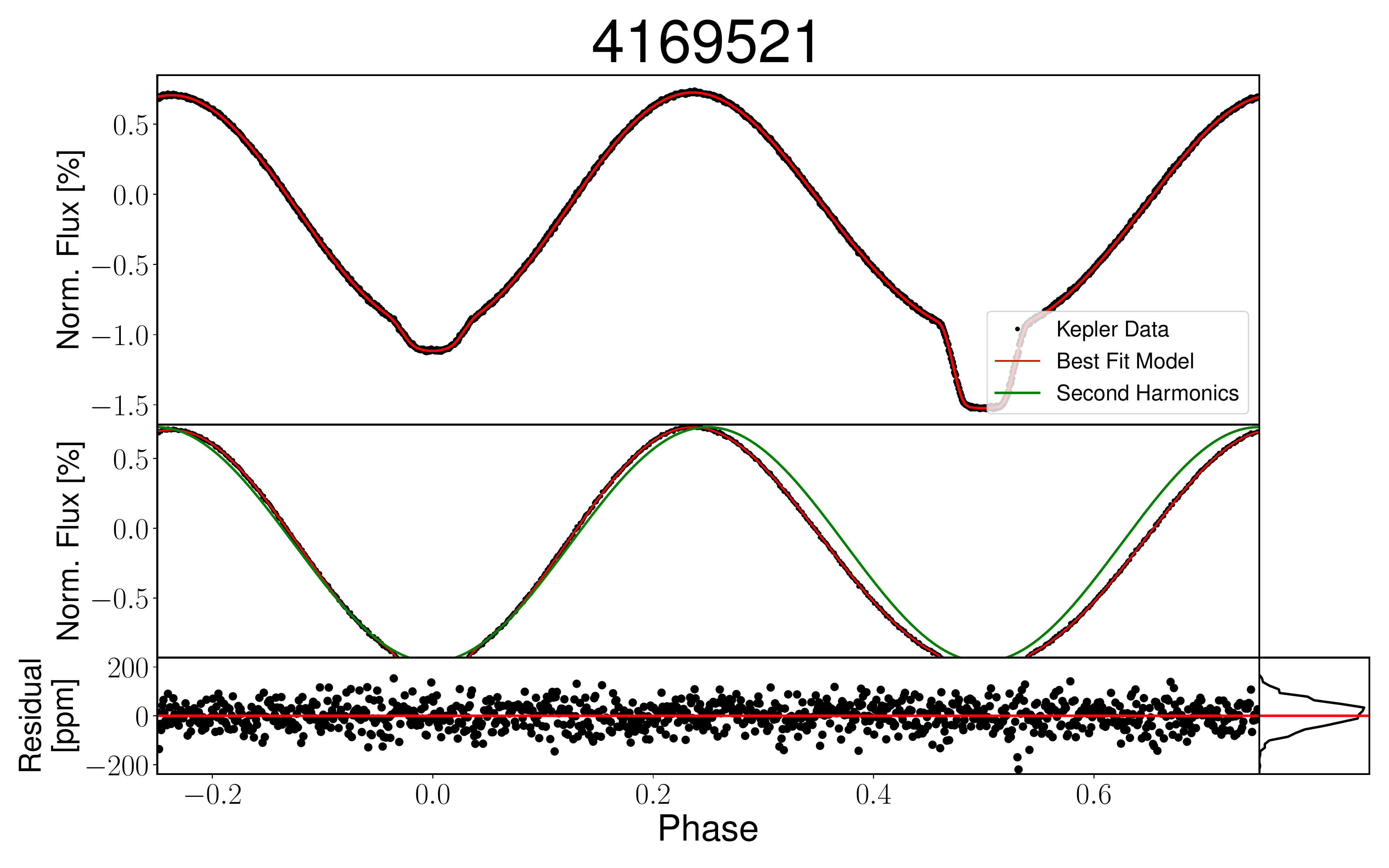} &
    \includegraphics[width=.315\textwidth]{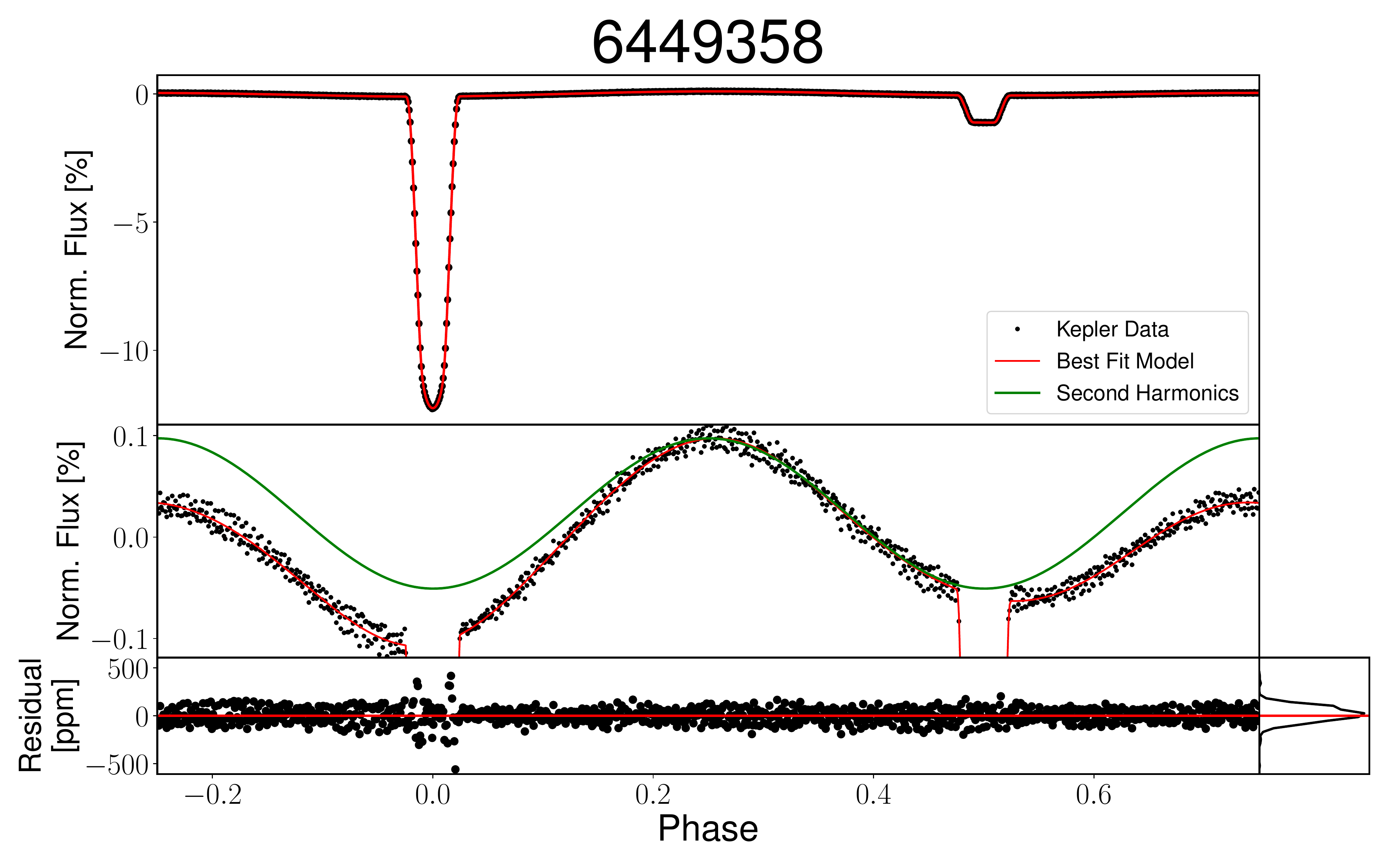}&
    \includegraphics[width=.315\textwidth]{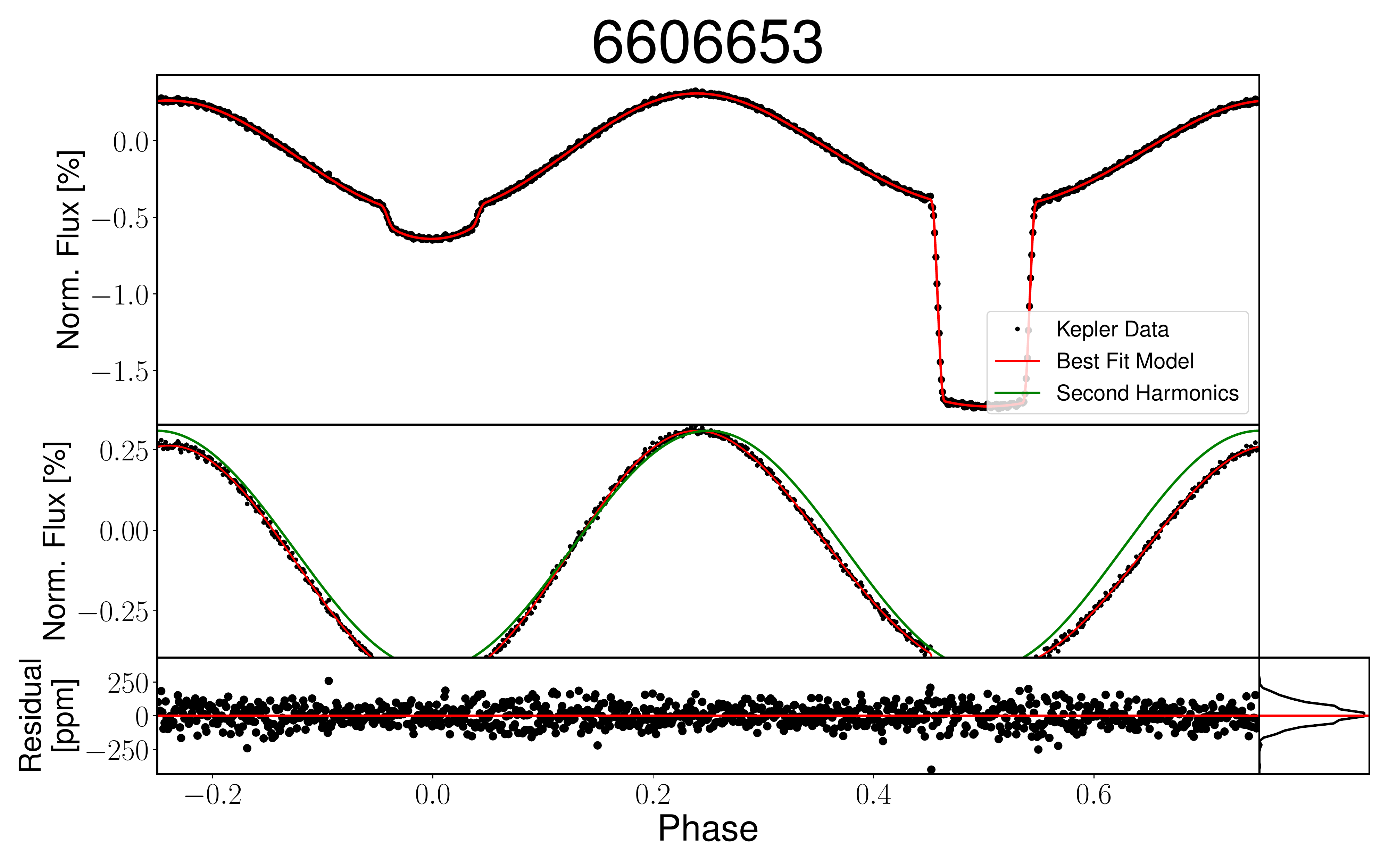}\\
    \includegraphics[width=.315\textwidth]{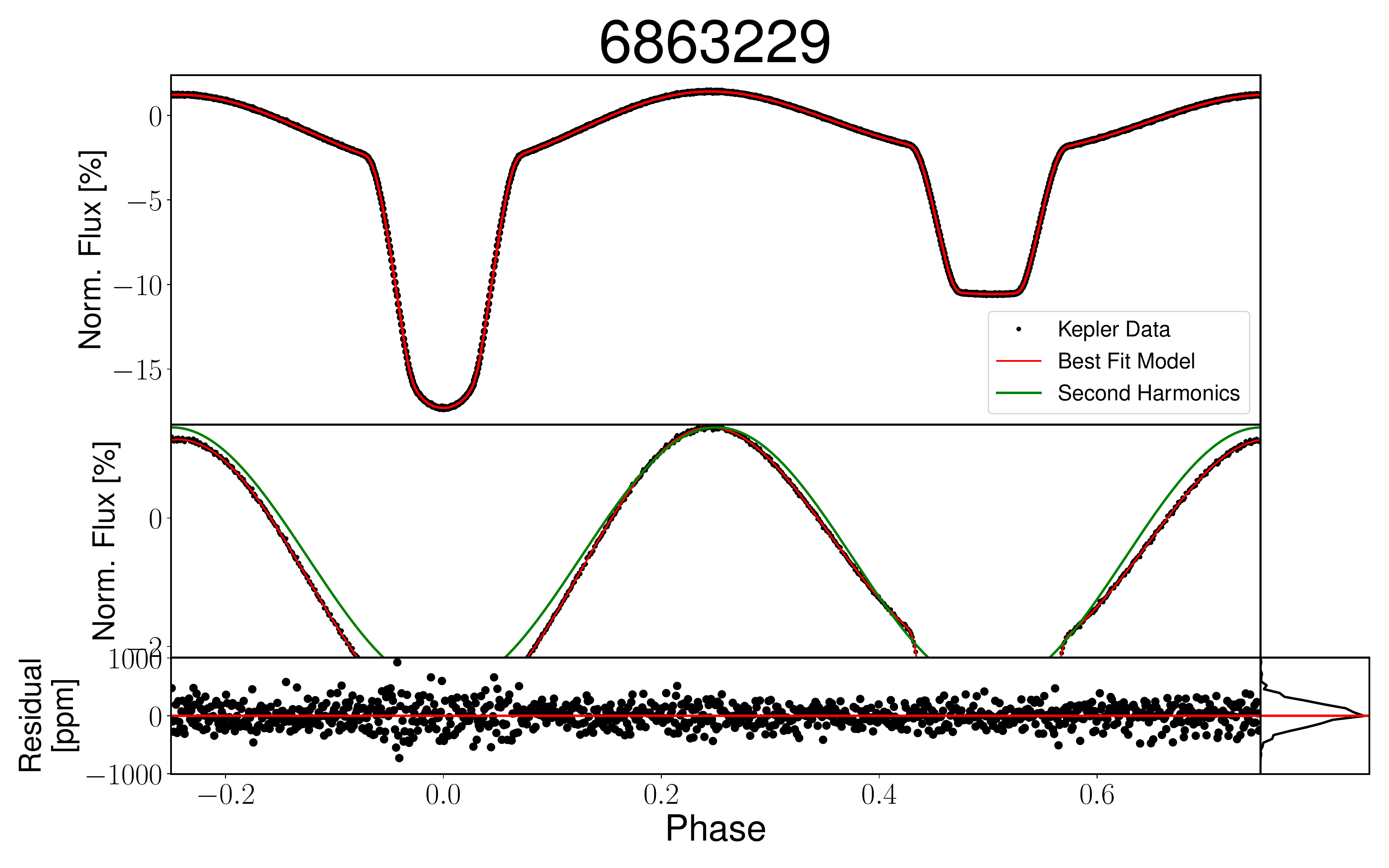}&
    \includegraphics[width=.315\textwidth]{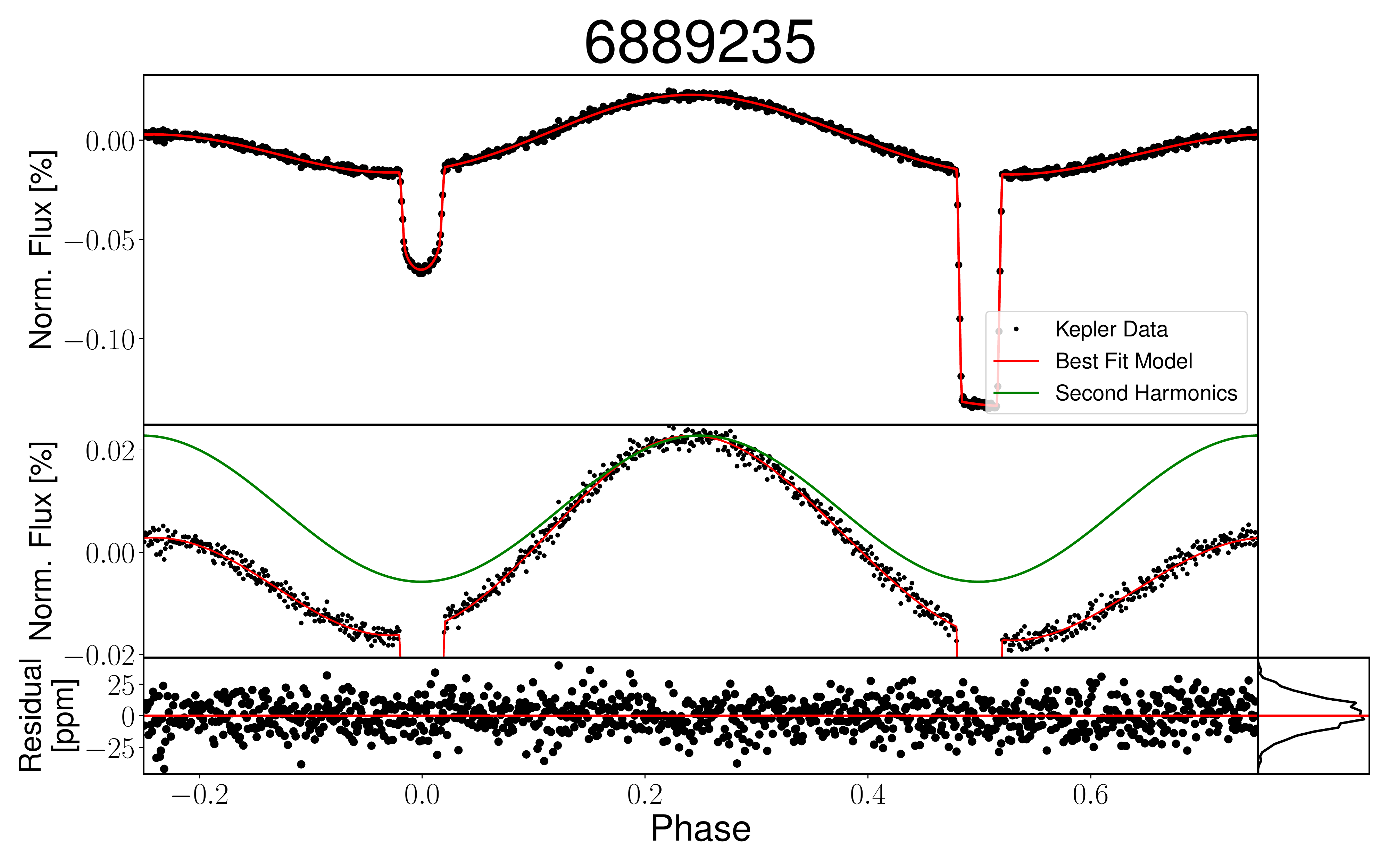}&
    \includegraphics[width=.315\textwidth]{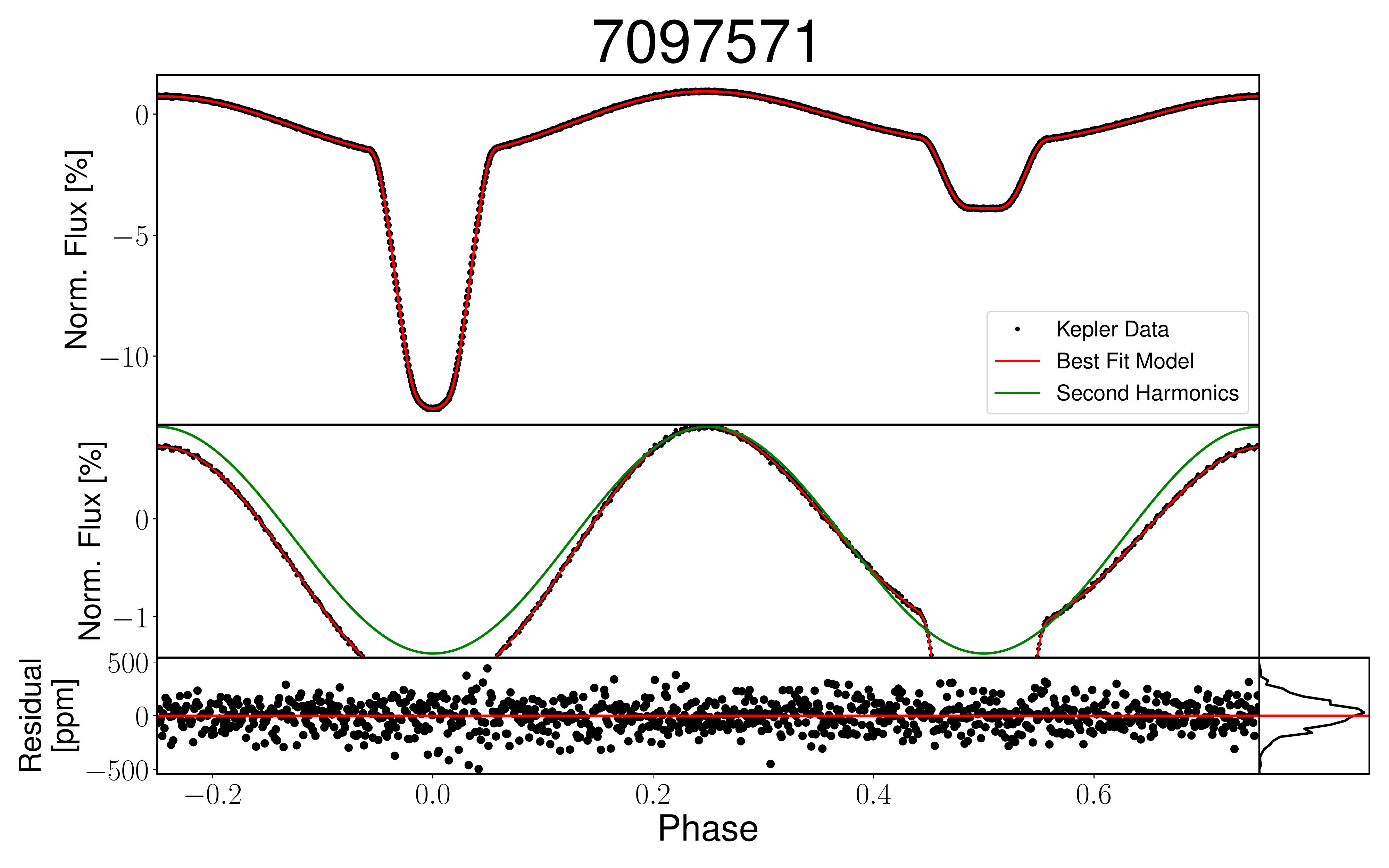}\\
    \includegraphics[width=.315\textwidth]{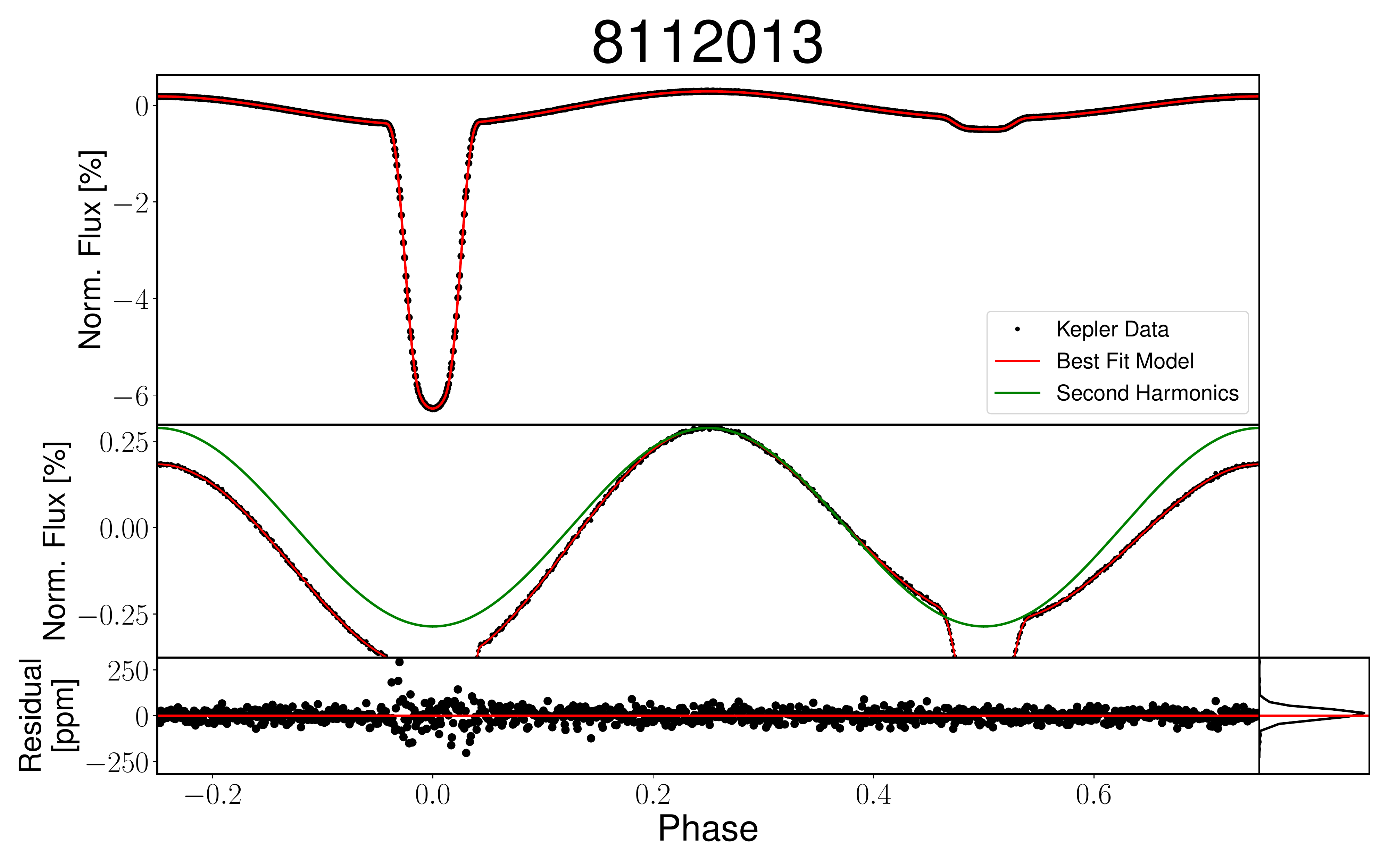}&
    \includegraphics[width=.315\textwidth]{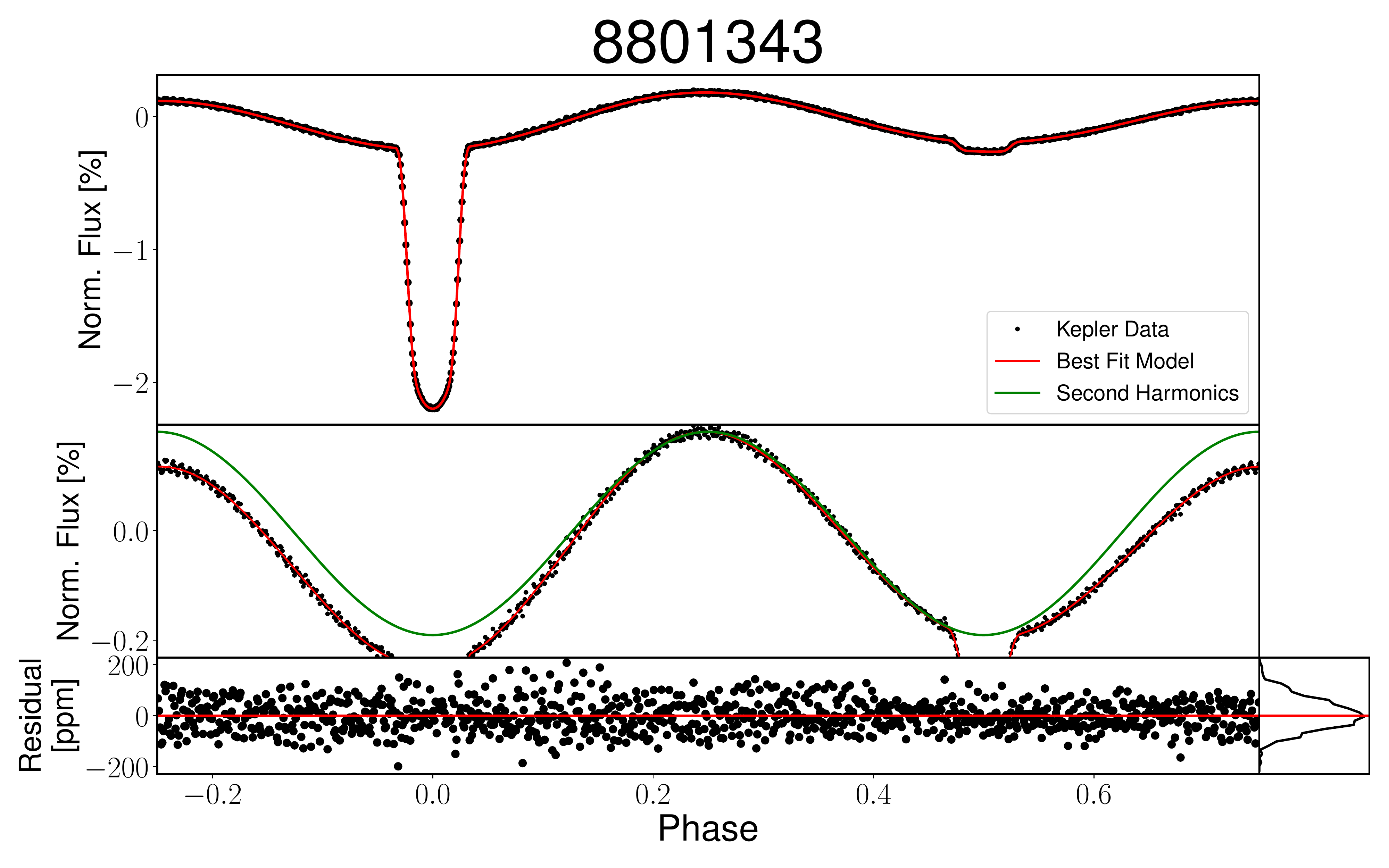}&
    \includegraphics[width=.315\textwidth]{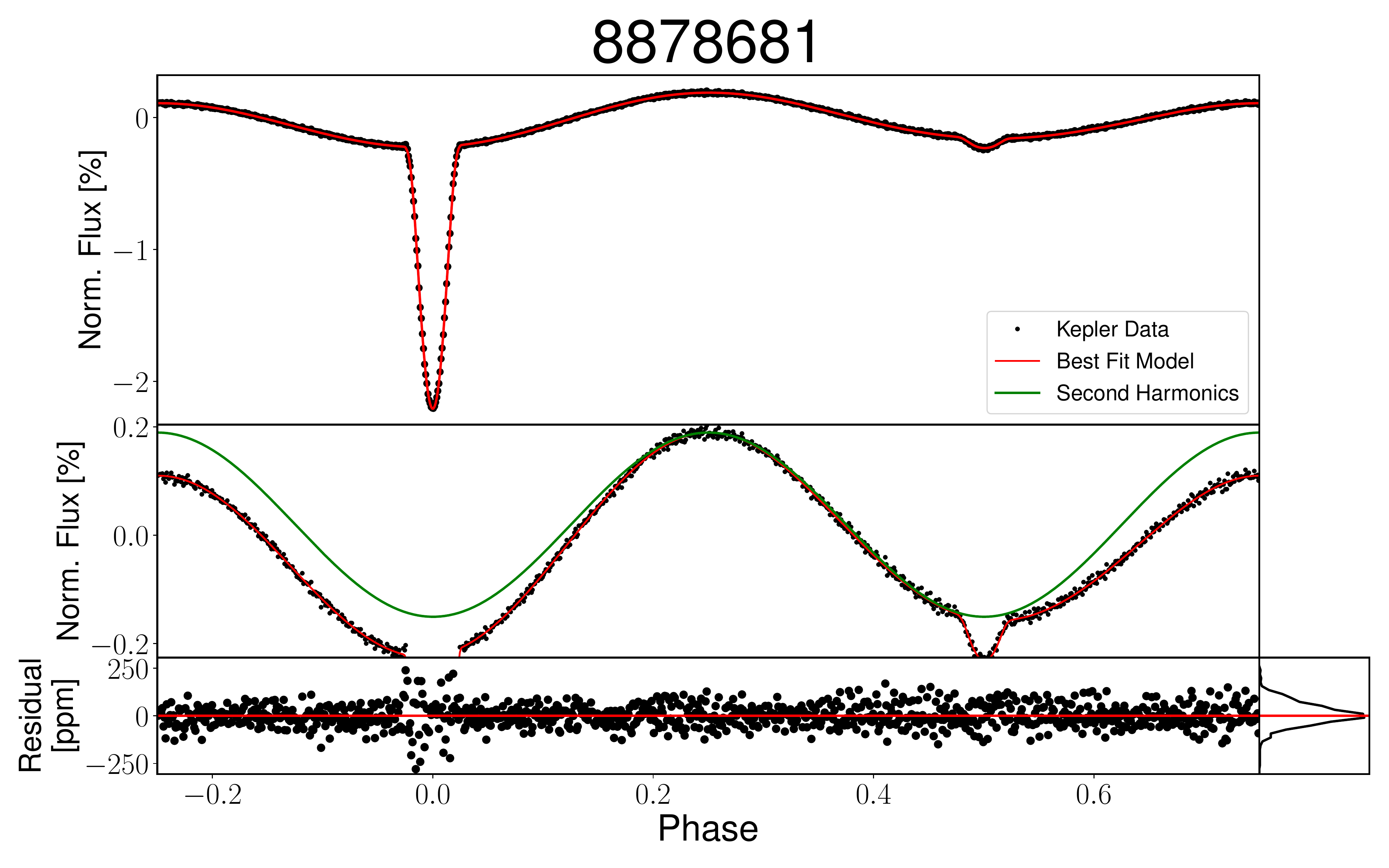}\\
\end{tabular}
\caption{Kepler EB phase-folded light curves (black) with best-fit models overplotted (red). Each panel shows the full phased light curve (top panel), a zoom in on the phase-curve modulation (mid panel), and the residuals from the best-fit model (bottom panel).}
\label{fig:BestFit1}    
\end{figure*}

\begin{figure*}[htb]
\centering
    \begin{tabular}{@{}ccc@{}}
    \includegraphics[width=.315\textwidth]{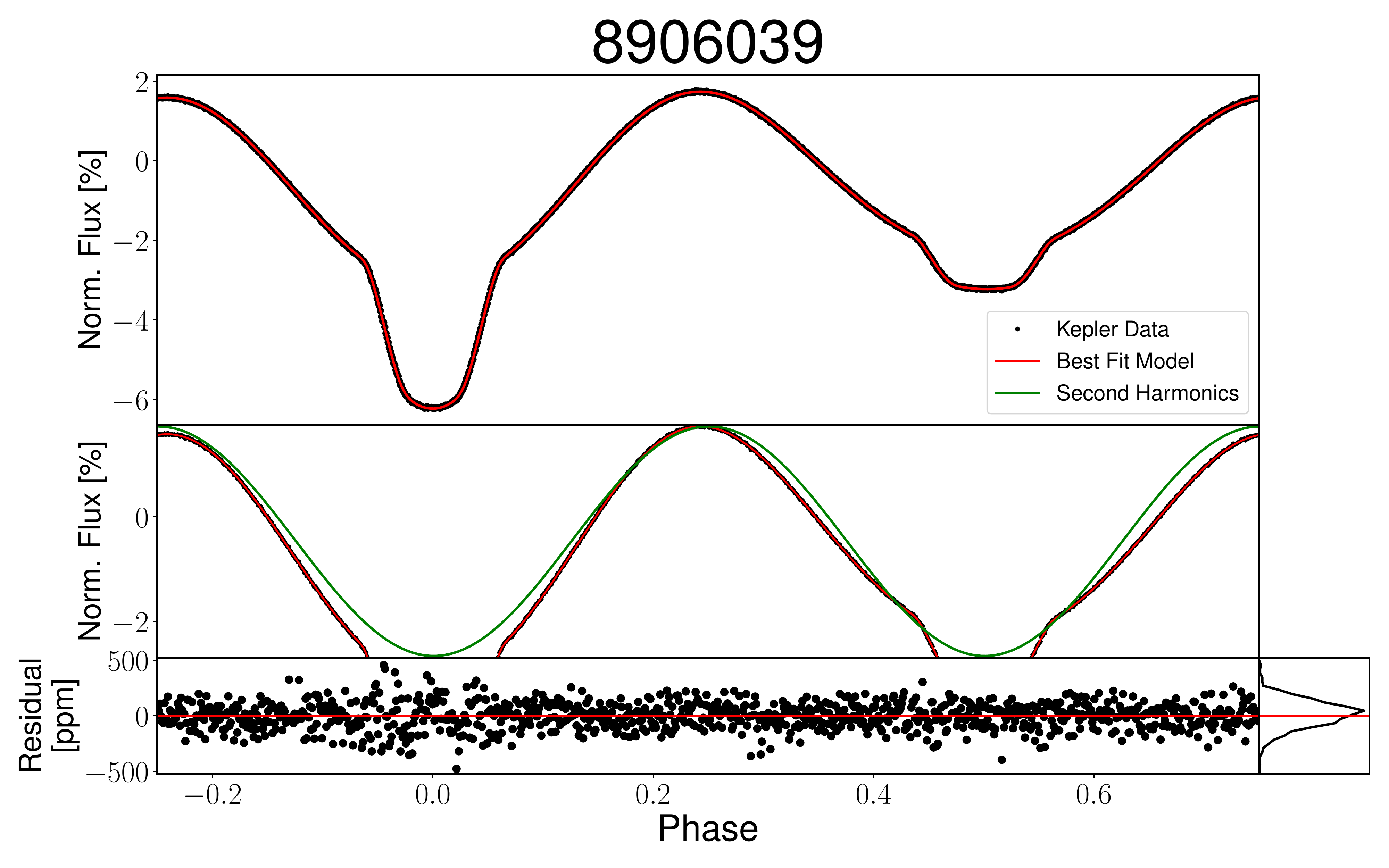}&
    \includegraphics[width=.315\textwidth]{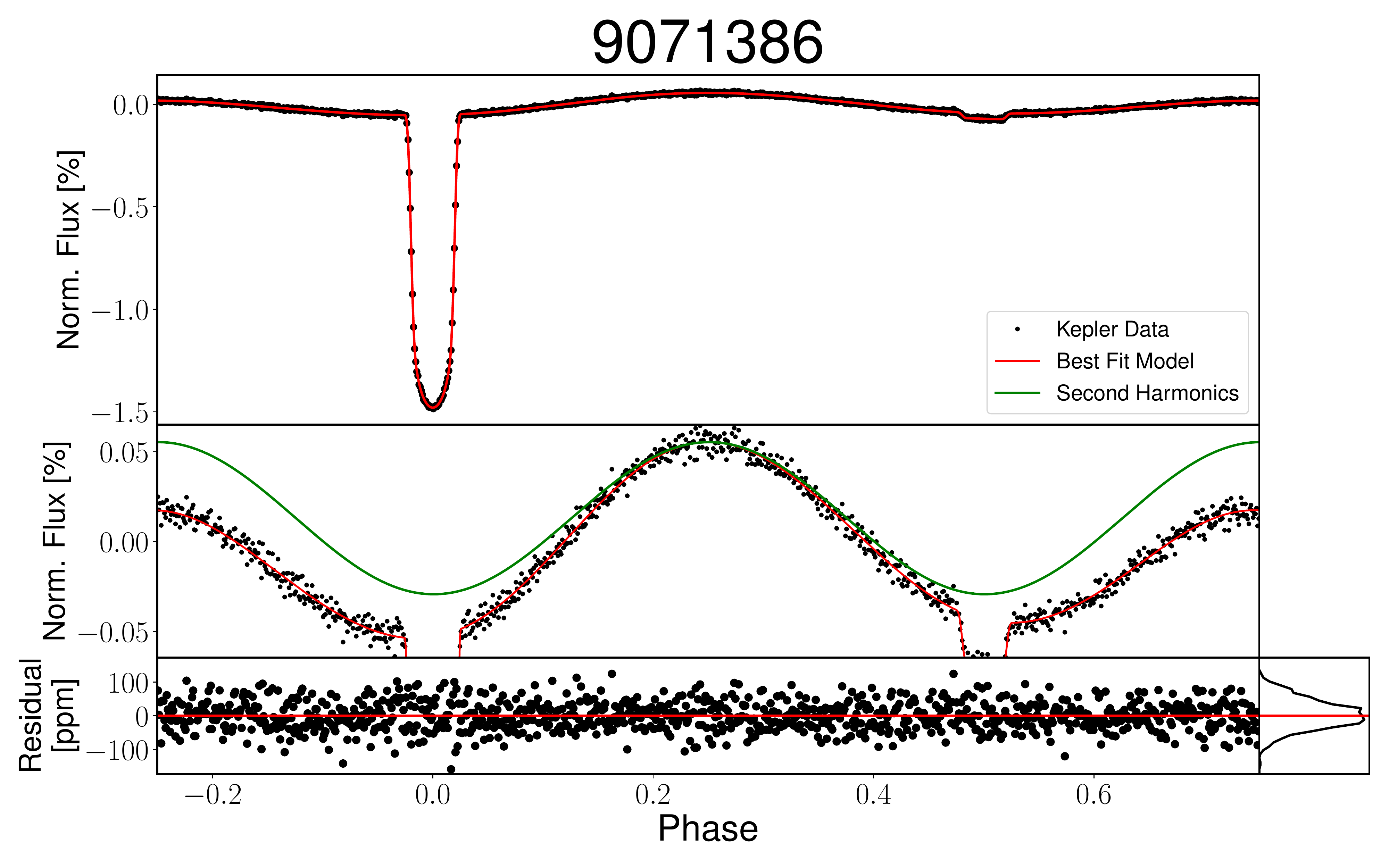}&
    \includegraphics[width=.315\textwidth]{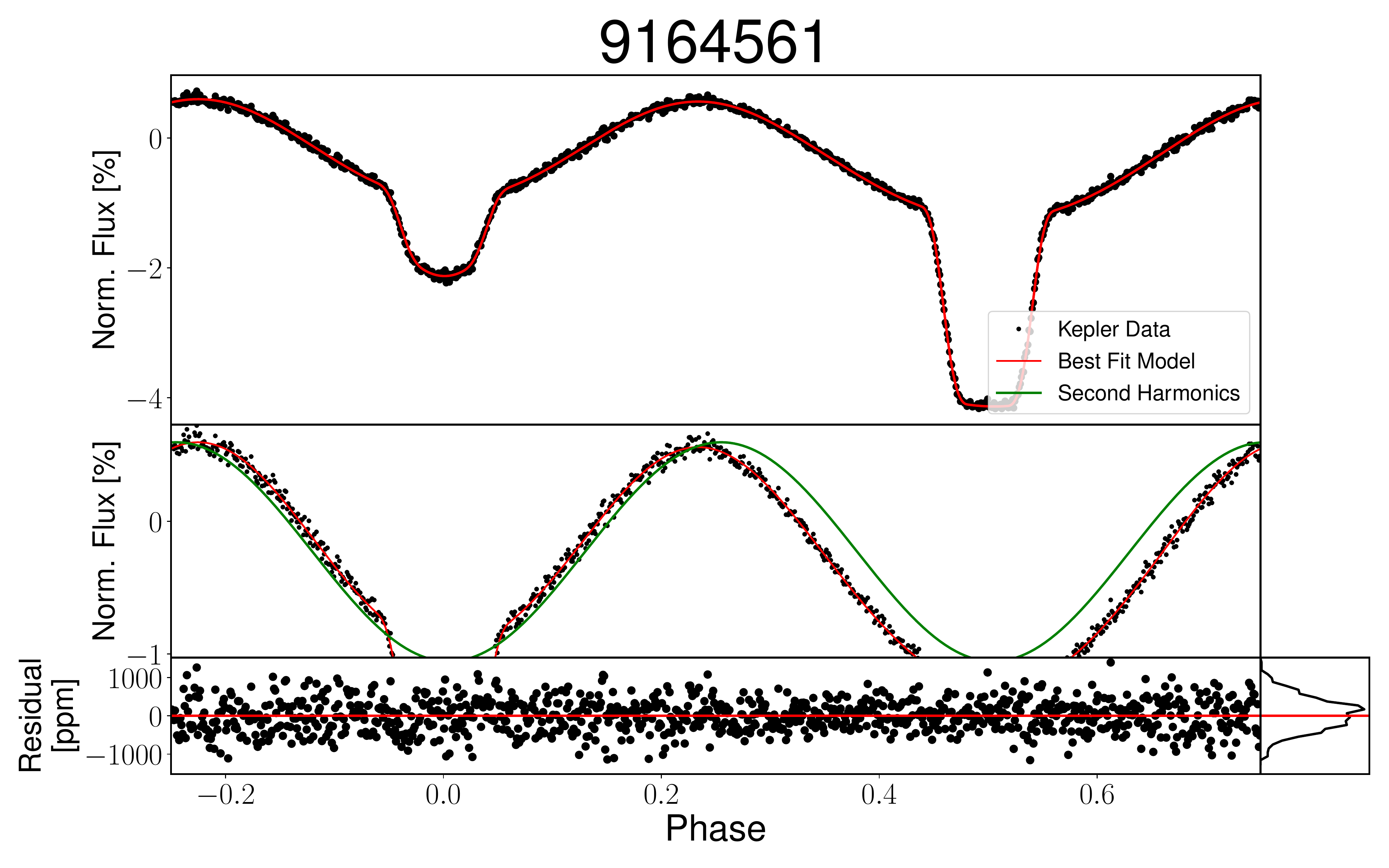}\\
    \includegraphics[width=.315\textwidth]{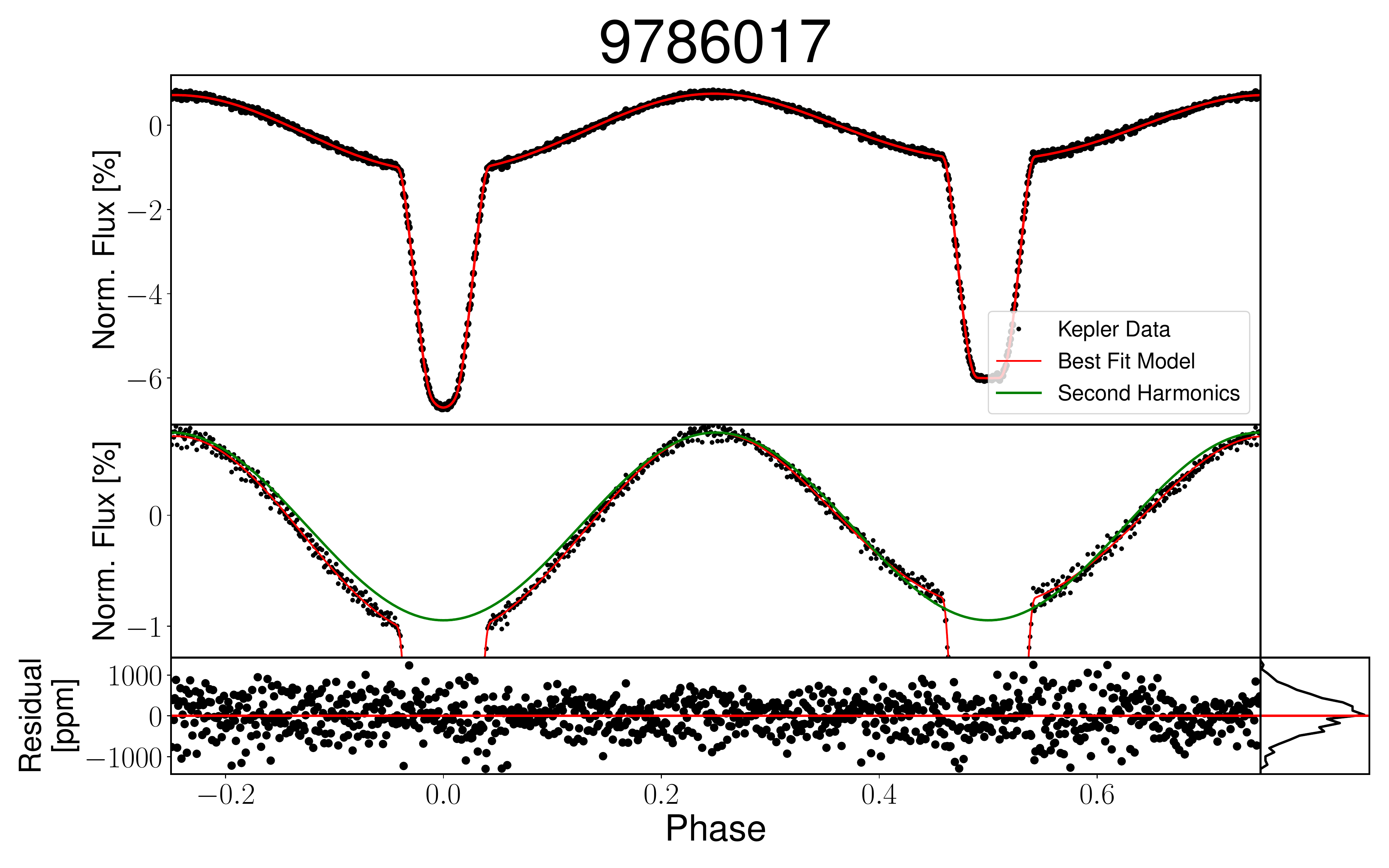}&
    \includegraphics[width=.30\textwidth]{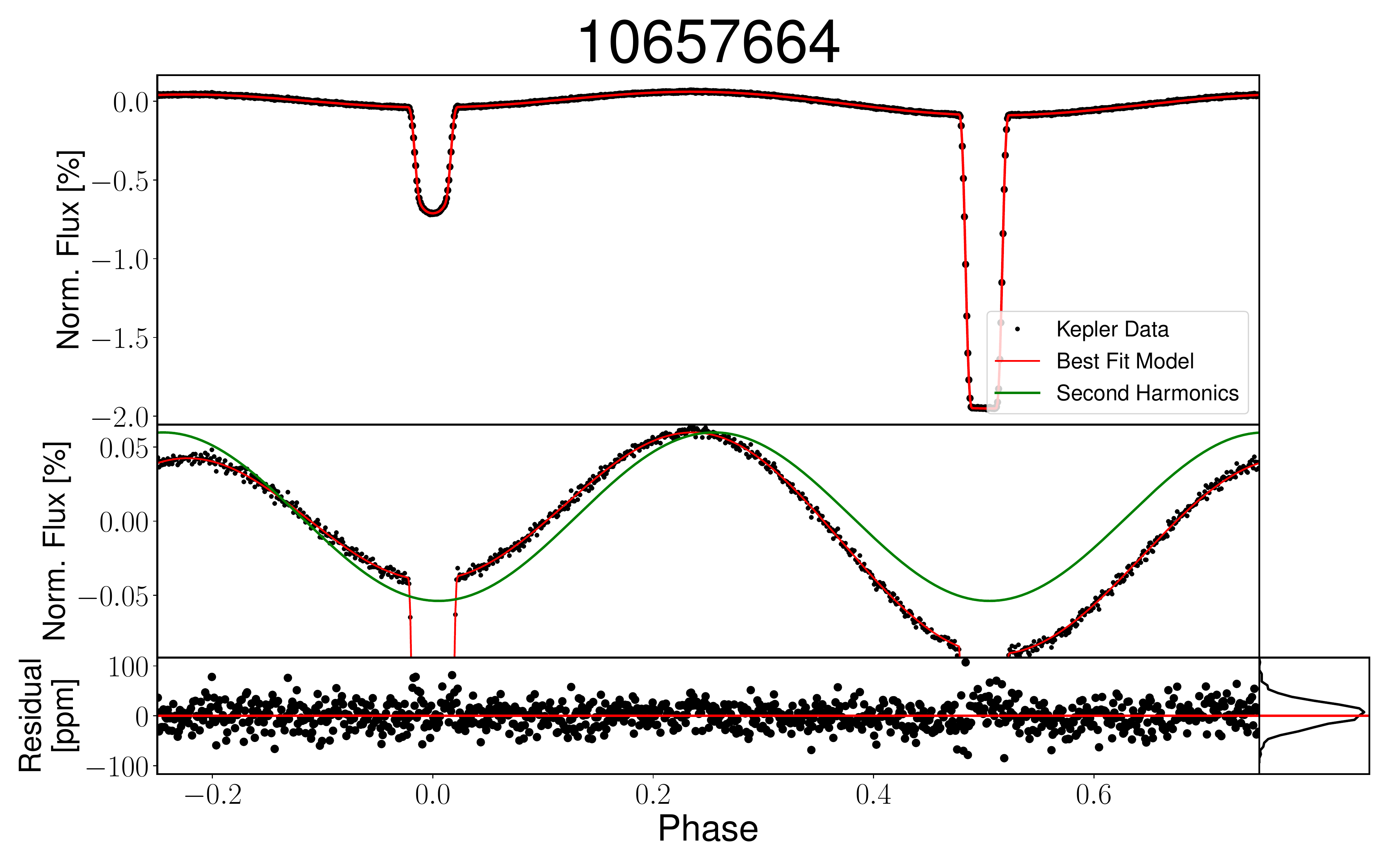}&
    \includegraphics[width=.30\textwidth]{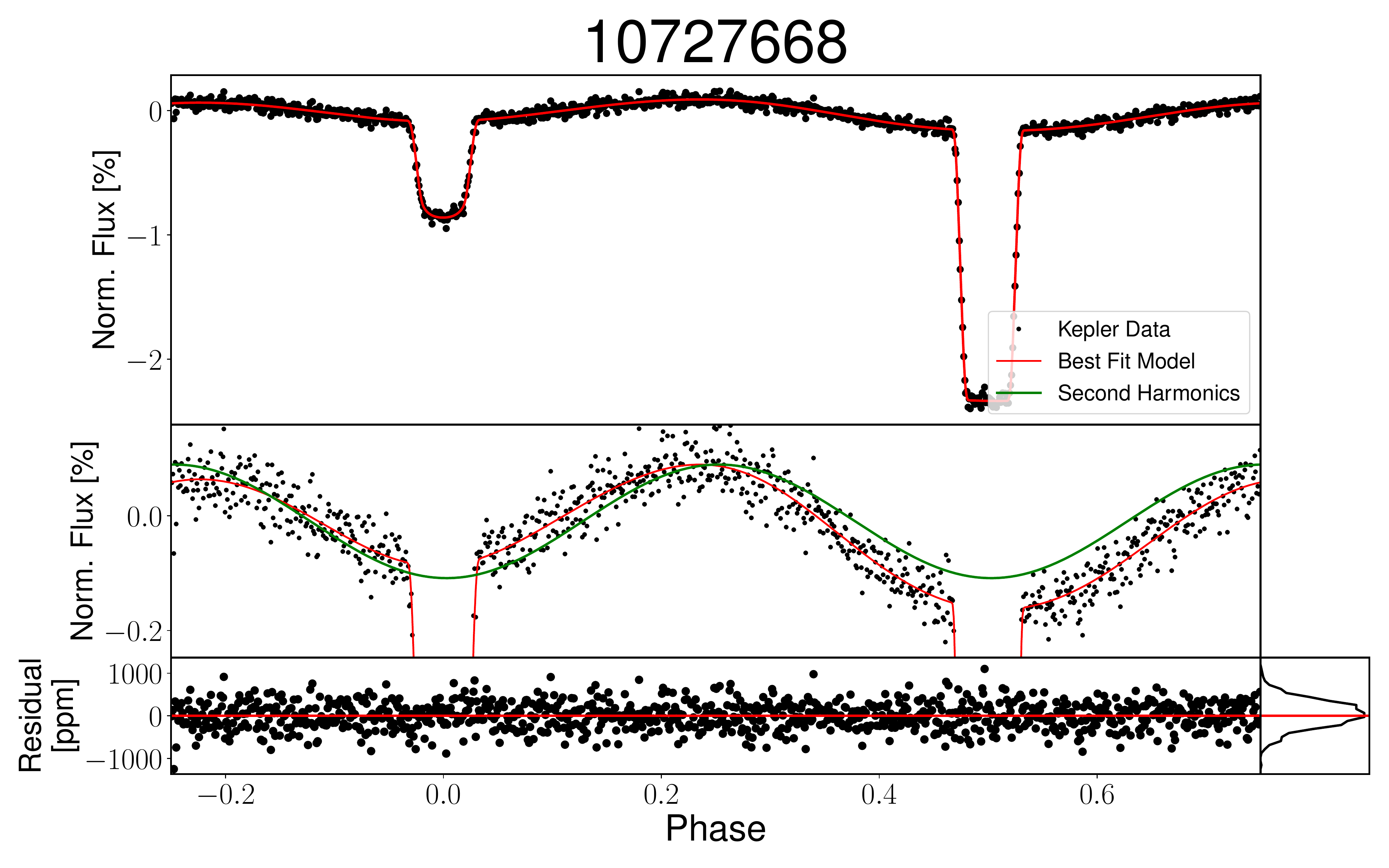}\\ 
    \includegraphics[width=.30\textwidth]{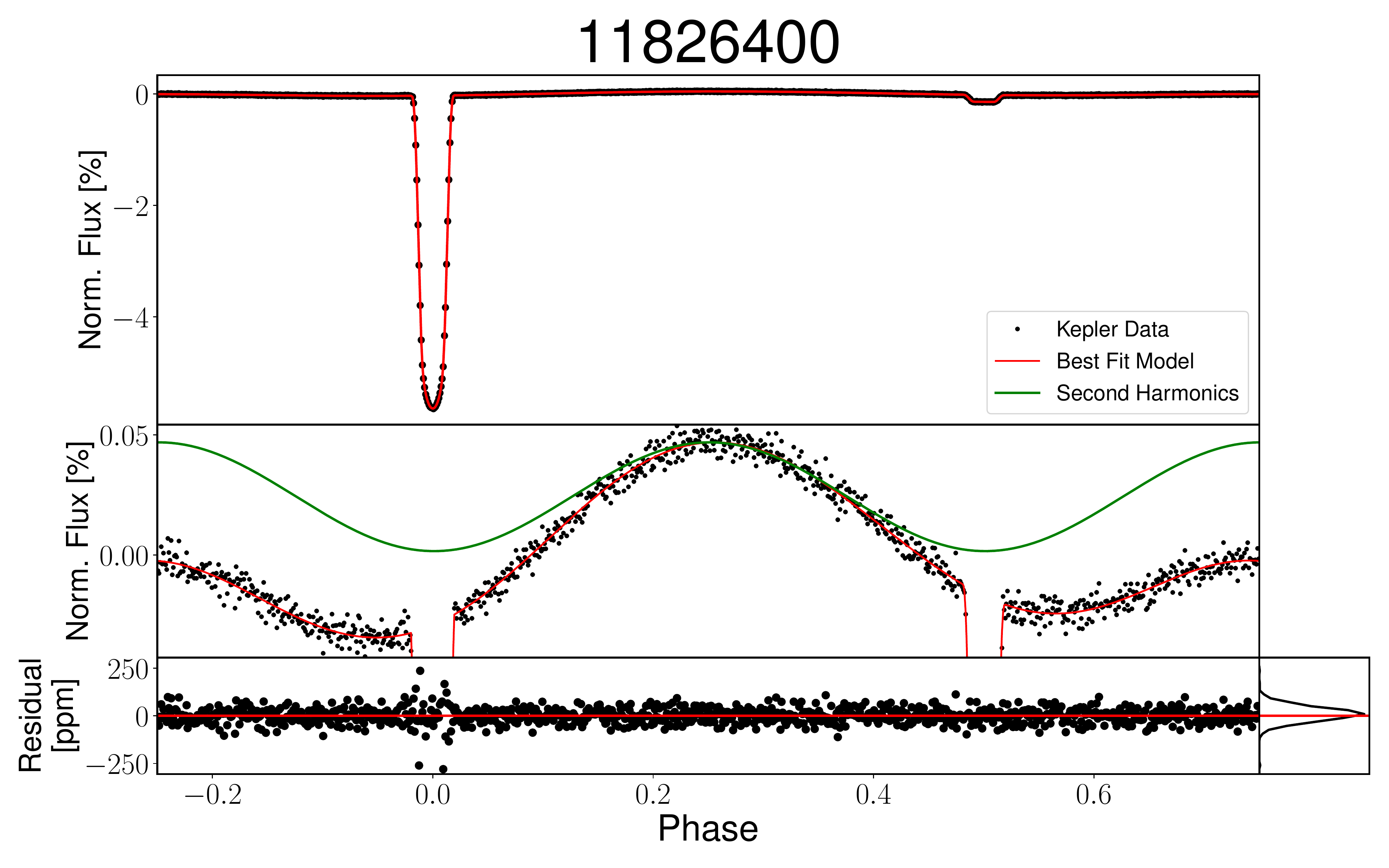}&
    \includegraphics[width=.30\textwidth]{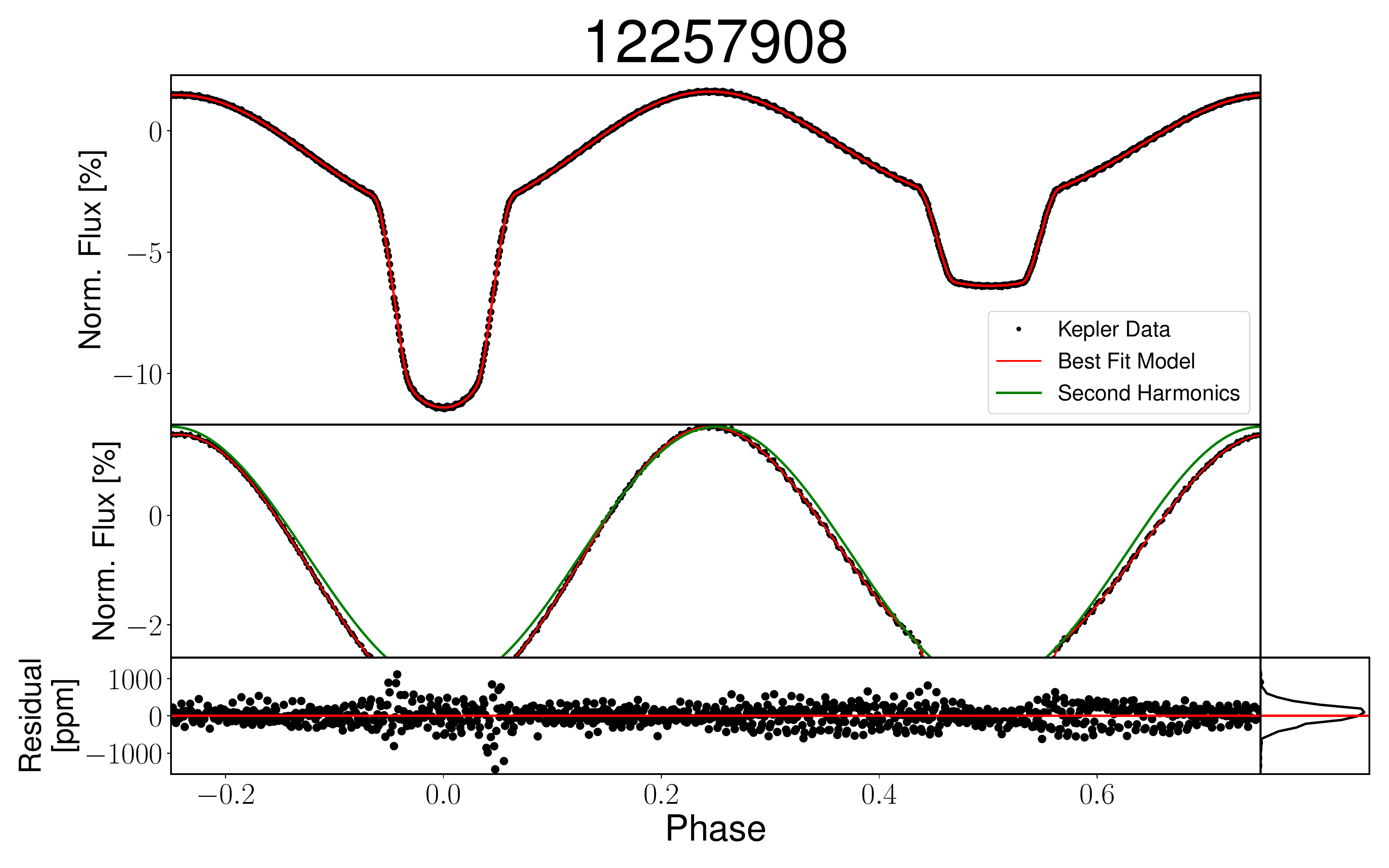}\\
\end{tabular}
\caption{Continuation of Figure~\ref{fig:BestFit1}.}
\label{fig:BestFit2}
\end{figure*}

\bibliography{EclipsingBinary}{}

\end{document}